\documentclass[sigconf]{acmart}

\usepackage{soul}
\usepackage[noorphans,vskip=1ex]{quoting}
\usepackage{wrapfig,lipsum,booktabs}
\usepackage[english]{babel}
\usepackage{blindtext}

\newcommand{\rev}[1]{\textcolor{black}{#1}}

\newcommand{\britta}[1]{\textcolor{orange}{}}
\newcommand{\ana}[1]{\textcolor{blue}{}}

\AtBeginDocument{%
  \providecommand\BibTeX{{%
    \normalfont B\kern-0.5em{\scshape i\kern-0.25em b}\kern-0.8em\TeX}}}

\setcopyright{acmcopyright}
\copyrightyear{2021}
\acmYear{2021}
\acmDOI{10.1145/1122445.1122456}

\acmConference[CHI '21]{CHI '21: ACM Conference on Human Factors in Computing Systems}{May 08--13, 2021}{Yokohama, Japan}
\acmBooktitle{CHI '21: ACM Conference on Human Factors in Computing Systems,
  May 08--13, 2021, Yokohama, Japan}
\acmPrice{15.00}
\acmISBN{978-1-4503-XXXX-X/18/06}

\acmSubmissionID{3331}

\begin{document}

\title{Fits and Starts: Enterprise Use of AutoML\\and the Role of Humans in the Loop}


\author{Anamaria Crisan}
\affiliation{\institution{Tableau Research}}
\email{acrisan@tableau.com}

\author{Brittany Fiore-Gartland}
\affiliation{\institution{Tableau Software}}
\email{bfioregartland@tableau.com}

\renewcommand{\shortauthors}{Crisan and Fiore-Gartland, et al.}

\begin{abstract}
AutoML systems can speed up routine data science work and make machine learning available to those without expertise in statistics and computer science. These systems have gained traction in enterprise settings where pools of skilled data workers are limited. In this study, we conduct interviews with 29 individuals from organizations of different sizes to characterize how they currently use, or intend to use, AutoML systems in their data science work. Our investigation also captures how data visualization is used in conjunction with AutoML systems. Our findings identify three usage scenarios for AutoML that resulted in a framework summarizing the level of automation desired by data workers with different levels of expertise. We surfaced the tension between speed and human oversight and found that data visualization can do a poor job balancing the two. Our findings have implications for the design and implementation of human-in-the-loop visual analytics approaches.
\end{abstract}

\ccsdesc[500]{Human-centered computing~Empirical studies in HCI}

\keywords{Data Science, Automation, Machine Learning, Data Scientists}

\begin{teaserfigure}
    \centering
     \includegraphics[width=\textwidth]{"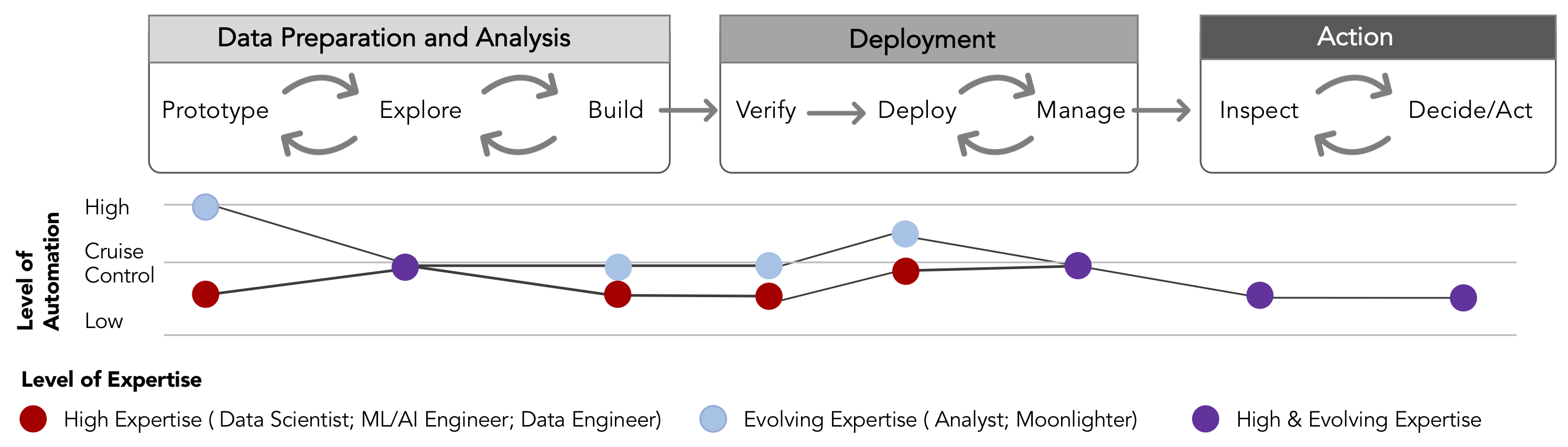"}
    \caption{Levels of Automation in Data Science Work.  From our interviews we illustrate the desired level of automation according to level of technical expertise in data science. We ground our findings in the levels of automation proposed by Parasuraman et al.~\cite{Parasuraman2000} and Lee et al.~\cite{Lee2019AHP} \\} 
    \Description{Illustration of the levels of automation (low, cruise control, high) across data preparation & analysis, deployment, and action. We also show steps for these high level processes. Data preparation and analysis contains prototype, explore, and build. Deployment contains verify, deploy, and manage. Finally, action includes inspect, decide/act. We show that the level of automation varies across the different processes, and also according to technical expertise - general individuals with low expertise required more automation than those with high expertise. However, individuals with high data science expertise still benefited from automation. }
    \label{fig:info_seek}
\end{teaserfigure}

\maketitle

\section{Introduction}\label{sec:intro}
Organizations are flush with data but bereft of individuals with the technical expertise required to transform these data into actionable insights~\cite{datasci_2017}. To bridge this gap, organizations are increasingly turning toward automation in data science work beginning with the adoption of techniques that automate the creation of machine learning models~\cite{Drozal2020,Wang:2019}. However, the adoption of this technology into enterprise settings has not been seamless. Currently AutoML offerings have limitations in what they can flexibly support. End-to-end systems encompassing the full spectrum of data science work, from data preparation to communication, are not yet fully realized~\cite{zoller2019benchmark,Lee2019AHP}. Consequently, AutoML systems still require human intervention to be practically applicable~\cite{Wang:2019,Passi2018}. This mode of human-machine collaboration presents a number of challenges~\cite{Liao_2020,Amershi2019}, chief among them being the importance of balancing the speed afforded by AutoML with the agency of individuals to interpret, correct, and refine automatically generated models and results~\cite{Heer_2019}. Data visualization can play an important role in facilitating this human-machine collaborative process~\cite{Heer_2019,Wang:2019}, but there are few studies that examine if and how data visualization is used in real-world settings together with AutoML. To fill this gap, we conduct interviews with 29 individuals from organizations of different sizes and that extend across different domains to capture how they currently, or plan to, use AutoML to carry out data science work. We examine specifically if and how participants use data visualization as a way to integrate the human in the automation loop.  

Our investigation reveals that the practical use of AutoML technology in real world settings requires considerable human effort. This effort is complicated by the need to trade-off data work between individuals with different expertise, for example data scientists and business analysts. This trade-off is exacerbated by a data knowledge gap that participants believe AutoML technology is widening. While participants saw the value of data visualization as one way to facilitate human-in-the-loop interactions with AutoML tools, many still reported using visualization in a limited way. Participants found that creating quality visualizations for AutoML was often too difficult and time consuming and had the effect of slowing down automation often with limited benefit. Moreover, participants reported a lack of useful visualization tools to support them in some of their more pressing needs, such as collaborating on data work among their diverse teams and with AutoML technology.  Altogether our study makes the following timely contributions to the existing literature on AutoML and the design of human-in-the-loop tools for data science: 

\begin{itemize}
    \item An interview study that presents real world uses of AutoML technologies in enterprise settings with a focus on the role of the human-in-the-loop facilitated by data visualization
    \item A summary of three use cases for AutoML according to different organizational needs
    \item A framework that illustrates the level of automation that is desirable for individuals with different levels of technical expertise.
\end{itemize}

As AutoML systems continue to gain traction in enterprise settings,  our contributions will be a resource to the research communities developing human-in-the-loop approaches that support an appropriate balance of automation and human agency.

\section{Related Work}\label{sec:related_work}
We review prior work that investigates the use of AutoML in data science, the ways that humans act within these processes, and current data visualization approaches that mediate these processes.

As we reviewed this work, we were challenged by the varied use of the term `AutoML'.  The preliminary goals of automation in machine learning began with the objective of removing the human specifically from hyper-parameter tuning and model selection steps~\cite{yao2018}. However, it quickly became clear that other steps, such as data preparation or feature engineering, were also critical to the success of hyper-parameter tuning. The scope of the term AutoML, and more recently ``AutoAI'' or ``driverless AI'', began to encompass broader steps in the data science workflow~\cite{Wang:2019,zoller2019benchmark}. We observed that the terms AutoML, AutoAI, and the phrase `automation in data science' are often used interchangeably in the literature. Here, we use the term AutoML to broadly encompass automation across multiple data science steps, from preparation to monitoring to deployment. 

\subsection{AutoML in Data Science}\label{subsec:bgpipeline}
Data science leverages techniques from machine learning to derive new and potentially actionable insights from real-world data~\cite{Donoho:2017,Longbing:2017,Blei2017}. AutoML systems have been developed to automate the computational work involved in building a data analysis pipeline that enable individuals to derive these insights from data. Several commercial systems already exists and are used within different types of organizations, including AWS SageMaker AutoPilot~\cite{sagemaker2020}, Google's Cloud AutoML~\cite{googleAutoML2020}, Microsoft's AutomatedML~\cite{azureaoutml2020}, IBM's AutoAI~\cite{ibm2020}, H20 Driverless AI~\cite{h202020}, and Data Robot~\cite{datarobot2020}.  There are also implementations of AutoML that build upon widely used data science packages, such as the scikit-learn~\cite{scikit-learn} python library, auto-sklearn~\cite{autosklearn2020,autosklearn2015} and TPOT~\cite{Olson2019,olson2016}. The focus of these AutoML systems are toward largely supervised tasks concerning feature engineering, hyper-parameter tuning, and model selection~\cite{zoller2019benchmark,yao2018,Manuel2014}. Recent innovations have proposed possible end-to-end solutions that also support data preparation~\cite{yao2018,zoller2019benchmark,Lee2019AHP} and it is likely that AutoML technologies will continue to expand toward broader end-to-end support. 

The means and extent to which AutoML systems integrate with a computational data science pipeline is variable. Some AutoML systems exist as a single component within a larger pipeline, such as automated feature selection step, that the analyst or data scientist creates. At other times, AutoML systems can also \textit{create} these pipelines with minimal user input. In their comprehensive analysis of existing AutoML tools, Z\"{o}ller et al.~\cite{zoller2019benchmark} describe three common configurations for including AutoML in data science work. The two configurations are "fixed structure pipelines'', where the AutoML system assumes a very specific configuration of computational pipeline. The authors differentiate between fixed pipelines that are optimized for specific AutoML methods (for example, neural networks or random forests) compared to those that are not. While these fixed systems are common, they have limitations when confronted with different data types and tasks. For example, image data or text data demand more flexibility within the structure of the computational pipeline.  The second category is a ``variable structure pipeline'', which refers to a fairly recent approach that aims to learn the appropriate steps within a data science pipeline~\cite{zoller2019benchmark}. TPOT~\cite{Olson2019,olson2016} is an example of one of the first variable pipelines. Unlike fixed models that \textit{execute} a pre-determined set of processes, variable structure approaches \textit{learn} a network of process in response to different datasets and user objectives.
\begin{figure*}
    \centering
    \includegraphics[width=\textwidth]{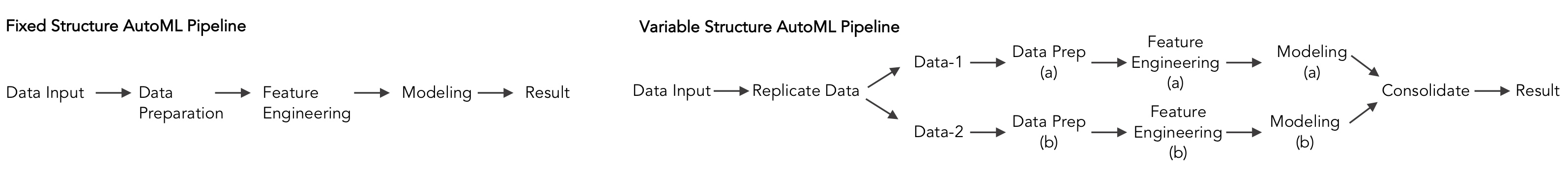}
    \caption{Example Illustrations of Fixed and Variable Structure Pipelines. Adapted with modification from Z\"{o}ller  et al.~\cite{zoller2019benchmark}\vspace{-5mm}}.
    \Description{All illustration of fixed and pipelines for AutoML. Fixed structure pipelines assume a concrete set of predefined step. The illustration show Data Input, Data Preparation, Feature Engineering, Modeling, and Result output. Variable structure pipelines do not assume a fixed set of steps. The illustration shows a Data Input step, and data replication step, and two parallel procedures that consist of preparation, feature engineering, and modeling, but that are intended to show different approaches to these steps. Next, the is a consolidation step to select the best pipeline configuration, followed by the presentation of the results. }
    \label{fig:pipelines}
\end{figure*}

While the stated goal of many of these AutoML systems is to effectively remove humans from many aspects of data science work~\cite{yao2018}, a view that data scientists themselves express~\cite{Wang:2019}, today these systems still rely on considerable human labor to be of use~\cite{Gray_Suri_2019}. These limitations stem from both the complexity of data science work and the brittleness of fixed structure pipelines that are in common use~\cite{zoller2019benchmark}. Our study catalogs this human labor across data science work and examines how visualization is used by individuals engaged in data work. 

\subsection{Automation and the Human-in-the-loop}\label{sec:related_work_hitl}
Human-in-the-loop approaches provide a way to explicitly incorporate human interaction within automated processes. Identifying when and how to add the human-in-the-loop within AutoML processes is important in order to appropriately balance the speed that automation affords with the importance of human guidance. Parasuraman et al.~\cite{Parasuraman2000} proposed a model to help designers identify the appropriate type and level of automation for information-seeking processes. They define four broad functions for how automation is used : 1) information acquisition; 2) information analysis; 3) decision and action selection; and 4) action implementation. They argue that the level of automation, from none to fully automated, should be evaluated against human performance consequences, automation reliability, and costs of actions. When the impact of automation is both significant and potentially harmful, human intervention is essential. The question of when, how, and how much to automate remains critical to the discussion of AutoML technologies today. 

A number of recent studies in the HCI literature have examined this trade-off between automation and human intervention as it relates to AutoML technology.

Lee et al.~\cite{Lee2019AHP}, Gil et al.~\cite{Gil2019} and Liao et al.~\cite{Liao_2020} describe a set of interaction modalities for users to engage with AutoML systems. Lee et al.~\cite{Lee2019AHP} categorizes Parasuraman's et al.~\cite{Parasuraman2000} levels of low to high automation into three different modes of interaction: `user-driven', `cruise-control', and `autopilot'. In `cruise control' a user directs an AutoML algorithm to a set of possible configurations to explore, as opposed to specifying a single and immediate next configuration. As an example, configuration can mean the user setting a parameter for hyper-parameter tuning during model creation. Gil et al.~\cite{Gil2019} describe a framework for human guided machine learning (HGML), which is predicated on the ability to effectively map user actions to a so-called `AutoML planner' capable of translating and executing the action. Similar to Gil e al., Liao et al.~\cite{Liao_2020} proposes a declarative way for the user to specify their objectives while allowing the system to automatically generate the underlying processes. By their descriptions, the systems proposed by Lee et al., Gil et al., and Liao et al. are akin to variable structure pipelines that were described in the previous section, in that they learn the processes in the pipeline. Studies have also examined human-ML/AI collaboration as it pertains to model authoring and interpretation specifically. While these studies are not exclusive to AutoML, they highlight key challenges for interacting with and interpreting machine learning models in enterprise settings. As an example, Hong et al.~\cite{Hong_2020} interviewed 20 individuals across different domains (the majority of whom identified as data scientists), and found that collaboration among different organizational roles was of chief importance for operationalizing machine learning models into organizational practices. 

Honeycutt et al.~\cite{honeycutt2020}, Liao et al.~\cite{Liao_2020}, and Amershi et al.~\cite{Amershi2019} describe the ways that information can be shared between humans and AutoML systems throughout a variety\britta{which interactions, interpretation?}\ana{modified sentence, but the description of the interactions follow} interactions. Honeycutt et al.~\cite{honeycutt2020} identifies `relevance feedback' and `incremental learning' as two general ways that humans can provide feedback to AutoML systems. Humans can provide relevant feedback, which informs the AutoML systems about whether its actions were effective or not. For example, humans may provide labeled data or correct errors when they arise. Humans may also provide new information in the form of incremental feedback to AutoML systems, which can be used to correct for issues like concept drift in models that have been deployed into production settings.  Liao et al. and Amershi et al. focus on the flow of information in the opposite direction, which concerns the types of information humans require to interpret the results of from AutoML systems. Liao~\cite{Liao_2020} conducted interviews with 20 UX design practitioners using a question bank to surface limitations in guidance targeting the development of explainable AutoML technologies. Their work demonstrates that the importance of the ML/AI results and their presentation is highly dependent on the question posed by the individual. Finally, Amershi et al.~\cite{Amershi2019} proposes a comprehensive set of 18 design guidelines that outline the appropriate modes of interaction when experts, a) initially interact with an AutoML system; b) as the system is churning; c) when errors surface; and d) throughout user interactions. 

Studies that examine how people use AutoML technologies and how they respond to human-in-the-loop features are also emerging. Wang et al.~\cite{Wang:2019} interviewed 20 data scientists across industries to interrogate their practices and perceptions of AutoML. They found the benefits of AutoML for augmenting, but not replacing, human intuition were valued and appreciated by practitioners. Passi et al.~\cite{Passi2018} conducted an extensive six month ethnographic study that involved over 50 data scientists. Their findings surface the different organizational needs and challenges of data workers as they collaborated with each other in the context of automation in data science work. Zhang et al.~\cite{ZhangY2020}, Drozal et al.~\cite{Drozal2020}, and Honeycutt et al.~\cite{honeycutt2020} conducted controlled experiments to evaluate decision making and trust in AutoML technologies, but their studies did not recruit current practitioners. Both Zhang et al. and Honeycutt et al. conducted their research via Mechanical Turk and Drozal et al. recruited undergraduate and graduate students in quantitative disciplines. Zhang et al. and Honeycutt et al. both found that reporting accuracy data alone was not sufficient for improving confidence and trust in the results produced by AutoML systems. Honeycutt et al. observed that the act of \textit{interacting} with a machine learning model reduced confidence of individuals in the model's performance even when the human guidance increased accuracy. These findings by Zhang et al. and Honeycutt et al. underscore the challenges of designing useful feedback mechanisms between humans and AutoML systems.  

While many human-in-the-loop approaches to support AutoML processes, and by extension data science work, exist, there are few studies aimed at understanding how they are integrated by practitioners in enterprise settings.  We found two studies that concretely explore AutoML in enterprise, and we build upon these findings in our present study in order to further assesses attitudes of individuals in enterprise settings toward human-in-the-loop approaches.

\subsection{Data Visualization and AutoML}\label{sec:related_work_vis}
Our work specifically focuses on visualization systems that support human-in-the-loop interactions for AutoML. Two prior and comprehensive state-of-the-art surveys capture the role of visualization in explaining~\cite{Chatzimparmpas2020} and building trust~\cite{ChatzimparmpasB_2020}  in machine learning. Recent work by Yuan~\cite{yuan2020survey} demonstrates visual analytic approaches throughout the data science process, including prior to model building (data prep and feature engineering), during model building, and after model building (verification, deployment). These surveys show the diversity of approaches that are taken to support decision making throughout the data science pipeline. Here, we highlight five systems that collectively capture this diversity. Google Vizier~\cite{googleVizer2017} and ATMSeer~\cite{wang2018atmseer} a) surface the complex latent space of models, b) search this space through interaction and visualization, and c) triage machine learning models. These systems present users with results from multiple models across their hyper-parameters through multiple coordinated views of the data. As with GoogleVizier, PipelineProfiler~\cite{pipelineprofiler2020} and AutoAIViz~\cite{Weidele_2020} make use of parallel coordinate plots to help users navigate the model search space and to highlight possible hyper-parameter settings. AutoAIViz shows the utility of conditional parallel coordinates plots to visualize subsequent steps in an AutoML pipeline based upon the user's current selections. One limitation of visualization for AutoML in data science pipelines is the assumption of a fixed structure (see Section~\ref{subsec:bgpipeline}), making it difficult to visually compare variable AutoML pipelines. To address this limitation, PipelineProfiler was developed as a wrapper for the auto-sklearn~\cite{autosklearn2020} package, supporting the visualization and comparisons of different end-to-end AutoML implementations.  

Taken together, we believe that these systems represent a `cruise-control' mode of including a human-in-the-loop, balancing between the slower `user-driven' and faster, but less transparent, `autopilot' modes for executing and interacting with AutoML. Moreover, these systems, co-created with experts in design and data science, represent real implementations of the existing design guidance toward the use of visualization to help interrogate AutoML systems.  However, it remains to be understood how such systems that are intended to build trust or transparency in AutoML actually get used, or perhaps more concerning, whether they get used at all. Our study sought to surface the visualization strategies within AutoML in enterprise settings. 

\subsection{Situating our Research}
The current state of the art in AutoML is informed by multidisciplinary research endeavours spanning machine learning, human computer interaction, and visualization. Given this research effort, there exist a number of AutoML offerings with varying types of pipeline configurations, from fixed to variable, and that support different modes of interaction so that ``intelligent services and users may collaborate efficiently to achieve the user's goals''~\cite{Horvitz_1999}. However, there remain few studies on how this technology is applied in enterprise settings, whether users can effectively leverage the benefits of this technology, and how adding the human-in-the-loop via visualization is viewed by enterprise users.  Moreover,  existing studies~\cite{Wang:2019,Passi2018,Hong_2020} looking at enterprise settings focused on specific themes, namely collaboration and trust, and did not closely examine how AutoML broadly intersects with data science work. Building on these prior findings, our study conducts a broader examination of AutoML and data science work that surfaces how AutoML is situated within organizational processes.





\section{Methodology}\label{sec:study_one}

\begin{table}[t]
    \centering
    \includegraphics[width=\linewidth]{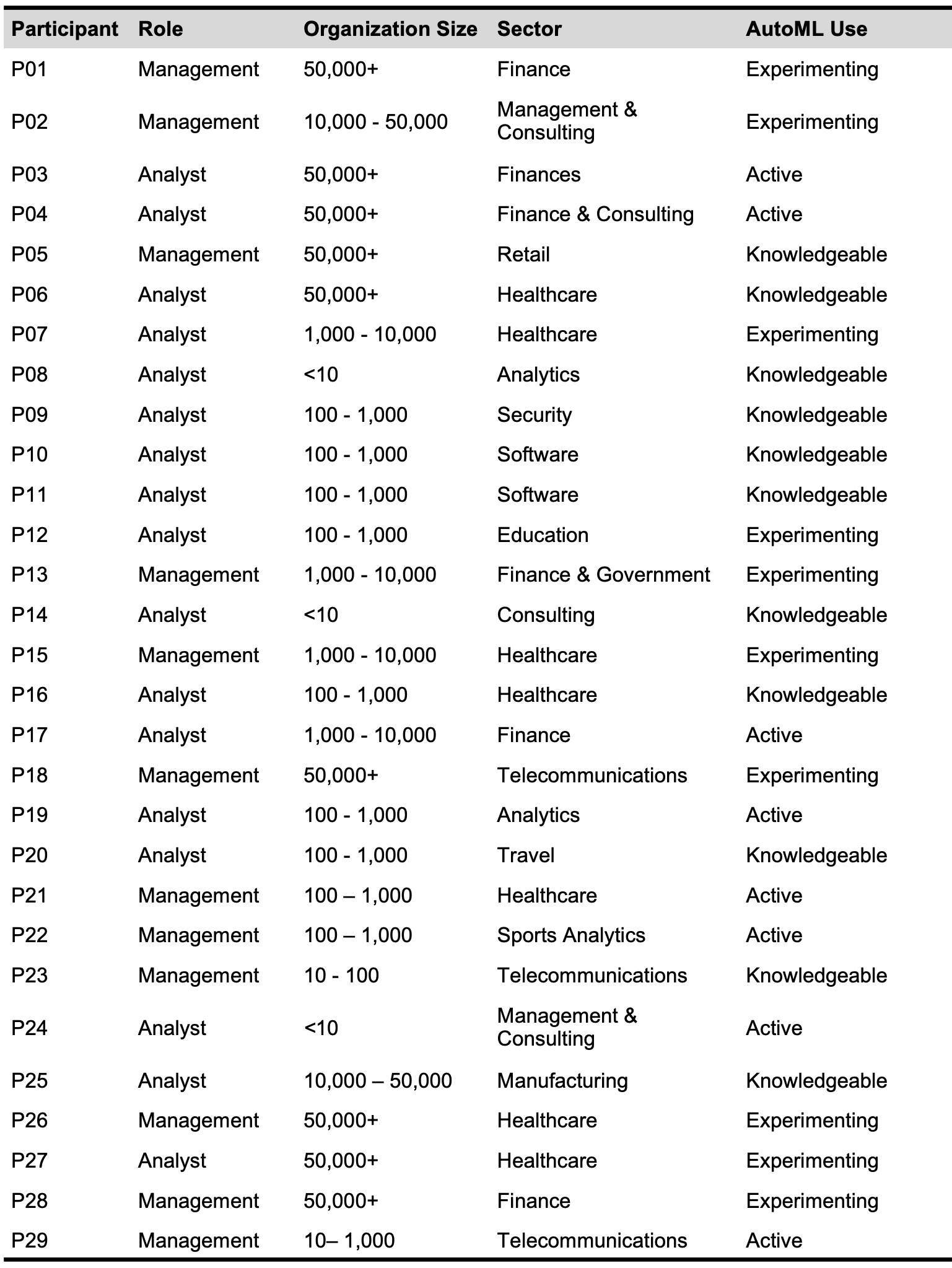}
    \caption{\rev{Summary of the Participants.}}
    \Description{A table with five columns that with information of the study participants. The five columns are participant ID, Role, Organization Size, Sector, and AutoML Use. The summary statistics for these columns are described in the text. }
    \label{tab:participant_table}
\end{table}

We conducted semi-structured interviews to develop an understanding of how AutoML is used to automate data science work. We were also interested in surfacing the role of the human-in-the-loop as it is mediated by data visualization tools, such as those used to explore data or support model tuning and selection. 

\subsection{Interviews and Data Collection}\label{subsec:study_one-interviews}
We recruited participants through a snowball sampling approach~\cite{Creswell_Poth_2018}, with the first point of contact being individuals that had participated in prior studies, were known to the authors, or other collaborators. \rev{We recruited and conducted interviews with 29 individuals that self-identified as data scientists, or analysts engaged in data science type work, or a manager overseeing a team comprising either entirely data scientists or a mixture of data scientists with others.}  The semi-structured interview format prompted participants to discuss data science work in their organization; if and how they currently use AutoML systems, or plan to deploy AutoML systems; and the ways they use data visualization, using the Tableau platform or other tools.

Interviews were scheduled for approximately 60 minutes, audio-recorded, and transcribed. \rev{Our participant screening questionnaires and interview guides are provided as supplemental materials.} Due to the nature of semi-structured interviews, the range of topics that participants chose to touch upon were quite broad. Moreover, due to the novelty and diversity of uses for AutoML technology the perceptions and pain points described by our participants were not always overlapping. All interviews were conducted over video conferencing software. A summary of participants, their organization size, and domain are summarized in~\autoref{tab:participant_table}.

\subsubsection{Sensitization to Emerging Concepts}
\rev{Sensitizing concepts are an important component of qualitative research, as they ground the analysis in important emergent features and operate as a key interpretative device in data analysis~\cite{Bowen}. At the outset of our study, we had some preliminary concepts that we were sensitized to from prior research we conducted that examined the nature of data science work and workers~\cite{Crisan_2021_dsframework}; we used this prior research as part of selective coding processes. In addition to this prior framework, we also had our own notions of concepts that could be pertinent to AutoML, visualization, and human-in-the-loop interactions specifically and these informed our initial interview questions.}

As we completed interviews we debriefed and conducted initial thematic coding of transcripts, we became sensitized to particular themes in our analysis that further refined our existing concepts of data science \britta{capitalizing?} work and generated new ideas that we had not previously considered. Specifically, these emerging themes included the importance of different types and levels of participant expertise and participants' use and attitudes around ``click'' (low or no-code solutions) as opposed to ``code-based'' solutions. Analysis of participant pain points surfaced issues of tool switching, trust, and collaboration. Finally, we also found that predictive modeling was the primary way that these organizations applied AutoML technology. As we became sensitized to these concepts, we revised our interview guide to ask more pointed questions about these themes. We provide both the preliminary and modified interview guides in our supplemental materials.

\subsubsection{Participant Characteristics and AutoML Use}
\rev{Participants self-identified as either analysts or managers. Analysts were individuals that were engaged in the day-to-day tasks of data analysis, including data scientists, business analysts, or other technical analysts engaged in data science work. Managers oversaw teams that often contained a mixture of data scientists, business analysts, or other types of organizational decision makers. In total, 17 participants in our study were analysts and 12 were managers. Overall, participants had high data science expertise, although one could be classified more as a citizen data scientist, which did not have formal training in data science but was exploring this field with the aide of AutoML.
Participants also represented organizations of different sizes performing a variety of functions. Four participants were at organizations with fewer than 100 individuals, 10 with between 100 to 1,000 individuals, 4 with between 1,000 and 10,000, 2 with between 10,000 and 50,000, and 9 with more than 50,000 individuals. Participants worked in a broad range of organizations across different industries that were focused on data analytics, finance, government, healthcare, management and consulting, security, telecommunications, and travel.}

\rev{We further stratified participants according to their current usage of AutoML technology. \textit{Active} users were those who reported that they, or members of their team, used a specific AutoML technology to conduct their work. We did not stipulate some required frequency of use (daily vs not) or the number of individuals currently using this technology. \textit{Experimenting} users were those that reported creating proof of concepts or described at least some preliminary projects specifically for the purposes of exploring AutoML technologies. Unlike active users, those that were experimenting with the technology articulated that their use of AutoML was in the early stages and exploratory in nature. Finally, those individuals that we categorized as \textit{knowledgeable} had high context for data science work including AutoML, but were not using or planning to use this technology in their work. Among our 29 participants 8 were active users, 10 were experimenting, and 11 were knowledgeable. }

\subsection{Selective Coding Process}\label{subsec:methods-analysis}
A prior study~\cite{Crisan_2021_dsframework} used an open coding process to define a framework of data science work that comprises  four \textbf{higher-order processes} and fourteen \textit{lower-order processes}. We use the set of codes from this prior study to carry out a selective coding of our interview transcripts. Selective coding is a stage in grounded theory research that serves to organize the analysis around a core set of variables~\cite{Bryant_Charmaz_2011}, in this case the processes of data science work, rather than derive and organize a new set of codes as is done in open and axial coding. The reason we use selective, as opposed to more commonly used open and axial coding approaches~\cite{Olson_Kellogg_2014} is due to AutoML technology being relatively new and as a result participants having varied experiences with it. While our interviews captured a rich diversity of experiences with AutoML this diversity also led to spareness in our data that made it difficult to achieve theoretical saturation in an open coding process. Using a selective coding process allowed us to scaffold our analysis around a cohesive narrative of how AutoML is used \textit{across} data science work. The selective coding process still makes use of constant comparison that allowed us to eventually achieve theoretical saturation in our findings. 

The set of four \textbf{higher-order processes} and fourteen \textit{lower-order processes} in this framework for data science work were:

\vspace{-0.5mm}
\begin{itemize}
    \setlength{\itemsep}{1.35pt}
    \setlength{\parskip}{1pt}
    \item \textbf{Preparation:} \textit{Defining Needs, Data Gathering, Data Creation, Profiling}, and \textit{Data Wrangling}
    \item \textbf{Analysis:} \textit{Experimentation, Exploration, Modeling, Verification}, and \textit{Interpretation}
    \item \textbf{Deployment:} \textit{Monitoring} and \textit{Refinement}
    \item \textbf{Communication:} \textit{Dissemination} and \textit{Documentation}
\end{itemize}

\vspace{-0.5mm}
The authors of~\cite{Crisan_2021_dsframework} also indicated two \textit{lower-order} processes, \textit{Collaboration} and \textit{Pedagogy},  that were identified as emergent but did not have sufficient context to place within the higher order categories. 

\begin{figure}[t]
    \centering
    \includegraphics[width=0.95\linewidth]{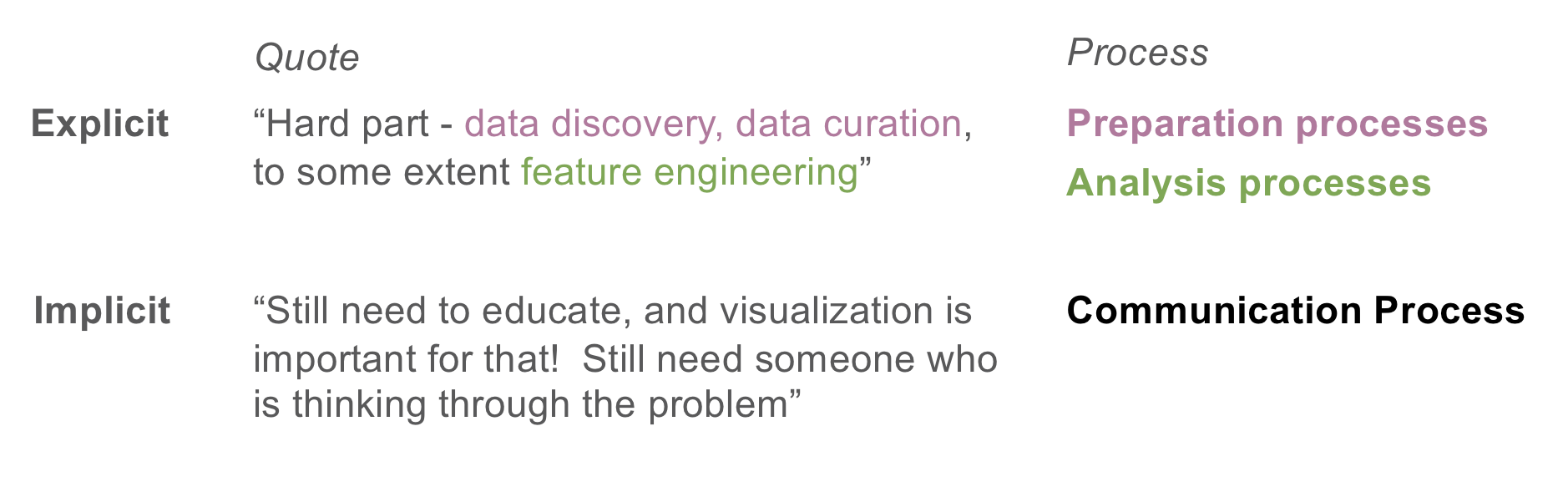}
    \caption{Example of Annotating Interviews with processes from an existing framework of data science work and workers~\cite{Crisan_2021_dsframework}}
    \Description{An example of how the data science framework from the previous study by Crisan et al. 2020 was used as part of selective coding processes. The diagram depicts the quotes and annotations that were described in the text.}
    \label{fig:annotation}
\end{figure}

In ~\autoref{fig:annotation}, we exemplify how we performed a selective coding  for these processes across our interviews. Some statements made explicit references to data science processes, for example, ``Hard part - data discovery, data curation'' are explicit references to the preparation higher-order process, as the terminology used can be linked directly to a higher order or lower order processes in the existing framework. By comparison, some references to processes were more implicit and were inferred by the authors with other context from the interviews. For example, ``still need to educate, and visualization is important for that. Still need someone who is thinking through the problem'' was determined to be an implicit reference to communication processes although there were not explicit terms specific to communication. 

As with putting any model into practice, the prior framework not only deepened our analysis, but also generated natural tensions between our observations and our framing of data science. We leveraged these tensions as points of inquiry that allowed us to critique or expand upon these frameworks based upon the participants' reported experiences. We reflect on our approach and propose modifications to it that we describe in Section~\ref{sec:framework-modifications} (as these modifications were motivated by our analysis) and again in discussion (Section~\ref{sec:discussion}).

\section{Results}
We first present our general findings regarding prevailing attitudes toward AutoML and the role of automation of data science work. Then, we examine the intersection of AutoML in data science work at a level of higher-order processes. 

\subsection{Attitudes Toward Automation}\label{sec:attitudes}
In this section, we describe prevailing attitudes toward adopting automation in data science as described by our participants.  We identified four primary themes that encapsulate these attitudes: the role of AutoML to drive productivity, the importance of tool integration, the concerns about automating bad decisions, and, finally, the desire to limit the role of the human-in-the-loop.

\subsubsection{Improving Productivity}~\label{subsec:attitude-productivity}
AutoML was embraced with cautious optimism by participants at organizations of different sizes, but we found that it tended to be more widely used or explored at larger organizations. It is not clear what is driving these differences, but we believe that it may relate to different amounts and rates of data collection at larger organizations that motivate a greater need for automation.  Among larger organizations that had implemented data science automation tools, the primary reason for investing in AutoML was that it ``just makes data scientists more productive''~(P01) by automating aspects of their work and allowing them to triage to focus on more pressing problems. As a specific example, P07 indicated that there is
\begin{quote}
    Lots of waste determining which models to work on.  [I] Wonder [if we] should focus on managing the pipeline so [that] the things get through have more impact. [We] need a predictive model to figure out which predictive models are the ones to work on.
\end{quote}

The automation of routine work, like model tuning and selection, was seen as a desirable way to shift the effort of human labor toward model verification and, if needed, correction tasks. While some participants felt strongly that technical expertise was required to safely use AutoML technology for productivity gains (a topic we return to in ~\ref{subsec:attidue-bad-decisions}), others (P03, P12) saw the benefit of AutoML to democratize data work. Individuals without a background in statistics and computer science that occupy roles of ``business analysts'' or ``moonlighters''~\cite{Kim:2017,Crisan_2021_dsframework} would benefit from the lower barrier to entry that AutoML affords. This democratization effort may improve productivity in those roles, but it also opens a door to capabilities across self-service analytics that were previously inaccessible.  

Overall, AutoML systems reduce the amount of code that is needed to use data within data science workflows. Individuals with high technical expertise, such as data scientists can leverage AutoML systems to improve speed and efficiency of routine tasks. For non-experts, AutoML systems can also democratize the accessibility of data science workflows and machine learning solutions. 

\subsubsection{The Importance of Integrating Tools}\label{subsec:tools_envs}
Participants reported using a variety of commercial tools to facilitate AutoML work and the need to create custom solutions for preparation, deployment, and communication data science processes. Participants used or were actively investigating platforms like Alteryx (n=5) for automating their workflows and integrating with Data Robot (n=3) or H2o.ai (n=3) to facilitate automatic modeling steps. Dataiku was also used in lieu of Alteryx and was seen as a better tool for facilitating collaboration across processes. Participants also reported using Sagemaker (n=2), PowerBI (n=2) (Azure), and Data Bricks (1). Participants report leveraging libraries like TensorFlow (n=4), scikit-learn(n=2), and mxnet (n=1) for their AutoML work. Python, R, and their attendant notebook environments were described as being used by roles that had higher technical expertise. However, P02 observed that ``businesses can’t deal with the notebooks'' because ``data scientist(s) are now [needing] to build things that can run in a production environment'' in order to operationalize models. Moreover, as organizations seek to spread out the data work  from data scientists to others in the organization, P15 observed that they were ``starting to see heavier reliance on data science products that don’t require heavy coding.''. For individuals without a data science or computer science background the need to write code appear to be a barrier, but our participants comments indicate that AutoML can lower, or potentially remove, this barrier.
Moreover, there is an increasing appetite for a ``platform [that] helps people move between tools that they have selected''(P04) and where  ``75\% of the organization could work''. Tool switching is common in data science because there are ``Different tools for different analysis'' and yet we found most users would ``prefer to stay in one environment''(P06).  Importantly, data science processes are not linear but occur in a ``big recursive'' (P09) loop \britta{is this a participant quote?}\ana{yes, added the participant info}. Constantly changing environments within multiple cycles of iteration and refinement is time consuming and impractical. 
Participants did report visualizing their data via systems like Tableau, or via charting libraries in R or Python, but many described their limited use:
\begin{quote}
As organizations scale, they're going to spend less and less time doing visualizations [...]  the job is to deliver results of a model in some form[...] Data scientists aren't going to deploy with ggplot, but they may use it for static reporting or just for their information. (P17)
\end{quote}

This participant didn't see visualization tools as scaling well alongside AutoML and other data science work, leading to abandonment, especially within results communication. Two other participants had to awkwardly move back and forth across modeling and visualization tools, and as a result the role of visualization was limited throughout the process.

\subsubsection{Concerns of Automating Bad Decisions}\label{subsec:attidue-bad-decisions}
Participants also clearly understood that blind trust in AutoML could lead to potentially catastrophic failures. P20 worries that ``lots of people will try to predict things without really understanding. [...] People will make horrific mistakes and not realize they've made them.'' P12 colorfully indicated that ``having this [AutoML] tooling may just allow people to make stupid mistakes easier!'' P05 observed that it's ``bad to slap together models and try to make decisions from it without understanding how things work. [That's like] giving a loaded gun to a child''. Participants were also concerned about regulatory constraints, for example the European GDPR legislation, that requires organizations to be able to explain decisions made by automated data science technologies. Even without legislative pressures, there were internal organizational concerns around these technologies, especially when large financial decisions were involved. A tolerance for errors and failure was an important factor in evaluating the use of AutoML technology; however, in many cases perfection was not required. For instance, P12 observed that there were ``lots of business use cases where 80\% accurate could be okay''. P29's observation echoes P12's that it is desirable for automation to help your surface failure points in your data preparation and analysis processes:
\begin{quote}
      I think they [executives] want you to use those [automated insights] to look at a graph and say, ``Oh wow, this is life changing. Let's go make this change in our business.'' We didn't use it like that. We used it to make sure that the results we were getting back made common sense. 
\end{quote}

A surprising finding was the general concern around the use of AutoML by ``citizen data scientists'' or domain experts that were not formally trained in data science, statistics or computer science. P12 stated that while they understand organizations want to democratize data science work, it still worries them because ``in practice you'll still have to be pretty technical'' to analyze data. P22 raised the issue of the overhead needed to ensure those ``who aren't as well versed...in the data science space are able to not make silly mistakes that they shouldn't be making''. Perhaps the strongest stance we heard was from a participant who stated that they would ``restrict it [AutoML] just to the data scientists'' and ``use it to get efficiency after demonstrating they know they are doing'' (P05). This attitude was a recurrent theme in our study. 

Overall, from participants' responses we see that the promise of automating data science is tempered by very real concerns of how things could go wrong. However, this does not mean that organizations are pulling back from their investment in these technologies. As P02 noted: ``ambition is still `industry 4.0' with lots of automation.''

\subsubsection{Limiting the Human-in-the-Loop}
Concerns toward safety and trust of AutoML in many ways highlight the value of humans-in-the-loop approaches to balance automation with human oversight. However, participants also expressed concerns about the inclusion of humans within the AutoML loop. For example, while P02 expressed that there was still ``lots of room for humans-in-the-loop'' innovations in their industry, they also stated that ``the manual part, where you have to visualize something, is getting cut out as much as possible''. P01 stated that even as "there are lots of automated tools in place making decisions'', at the same time there was a lot of ``anxiety in the firm about what people do with ML,'' stemming from concerns of automating decisions at scale. We interpret these seemingly contradictory positions to mean that human-in-the-loop approaches are valuable when applied at the right time and in the right way. 

\subsubsection{Summary}
\rev{We briefly summarize the key takeaways for participant attitudes toward AutoML. While participants expressed concerns about the potential to automating bad decision-making, there was also a growing  interest in using AutoML technology to produce a 'good enough' result that could scope out the viability and possible issues with the data or machine learning product. AutoML allows for the creation of sophisticated tools with minimal code and offers opportunities to `fail fast', which enables data scientists, and even so-called `citizen data scientists', to surface issues earlier.\britta{see point above}\ana{addressed}  However, what is most clear is that applying AutoML technology at suitable points in data science work is very important, otherwise it is dismissed as intrusive. To further explore \textit{when} and \textit{where} AutoML technologies could support data science, we analyzed interviews through the lens of an existing model for data science work.}

\subsection{AutoML in Data Science Work}\label{sec:auto_in_ds}
In this section we summarize our findings on the use of AutoML in the data science process. We consider places where considerable human labor is required either to support the creation of a machine learning model or to interpret, communicate, and act on its findings. Following the framework described in Section~\ref{subsec:methods-analysis}, we begin by examining AutoML in data preparation, followed by analysis, deployment, and communication. 

\subsubsection{Data Preparation Continues to be a Rate Limiting Step for Automation}
Participants understood that without robust data preparation the AutoML portion of their Data Science processes would be ineffective. P02 succinctly stated that,
\begin{quote}
    AutoML has never been the solution - it’s a shiny toy. [It's] always going to be about the quality of the data - do you understand what you are modeling?
\end{quote} 

In our interviews, participants identified several challenges in data preparation work that still required a lot of human labor, from gathering data, profiling it, and wrangling it into shape for analysis. It is important to emphasize that participants rarely began by cleaning a single tabular dataset, but often needed to bring several data sources together. P12 reported that ``40-50\% of my team’s time [was spent] on Alteryx to bring data together'', while P23 reflected that ``the most difficult part of my day is getting the data that I need to work with''. Once the data were gathered, participants faced difficulties with data profiling, a challenge that was also surfaced in prior work by Kandel~\cite{Kandel:2012} and Alspaugh~\cite{Alspaug:2018}. P19 expressed that automation in 
\begin{quote}
    Data profiling would be a huge win. I spend a ton of time having to explain the shape of data, and what shapes work best, how to explore the data, and how to refactor as needed
\end{quote}

Even when participants have data gathered and profiled, they still need to assess its utility for further downstream analysis. Rapid iteration via AutoML plays a role in speeding up this manual process. P23 described that they ``get [the data] to a point where maybe it's 60\% [clean], and they start to run algorithms...[in order to]...expand on the data itself''. The workflow described here resonates with rapid prototyping and failing fast to discover issues or limitations in the data. This observation also echoes data reconnaissance and task wrangling processes previous described in~\cite{2020_datarecon}, in which individuals acquire and quickly view data in order to assess its suitability for analysis and decide whether to pursue additional data sources.  Despite the usefulness of AutoML-driven prototyping, challenges associated with data preparation remain. This reflects an emerging theme from our analysis that there is a growing need to more tightly couple investments in data prep and model building. For instance, the experience of these challenges led P03 and their team to make more significant investments toward ``tools for democratization of data prep and (to a lesser extent) model building''. 

It is not surprising that data preparation is both time consuming and important to the successful application of AutoML and data science more generally. Prior studies~\cite{Kim:2017,Wang:2019} with data scientists and other domain experts have routinely pointed to this bottleneck for years, and visualization tools such as Trifacta (and its academic predecessors Wrangler~\cite{2011-wrangler} and Profiler~\cite{2012-profiler}), and Tableau Prep have been developed to address this challenge. It is disconcerting that preparation continues to be such a significant bottleneck despite existing tools. Our observations suggest that one reason for this may be that existing tools for data preparation do not easily fit within a data workers' analytics environment.~\rev{By extension of these observations, AutoML technology needs to be well-integrated with existing tooling environments, while also surfacing the manual labor and lack of adequate tooling for data preparation.}

\subsubsection{Use of AutoML in Analysis Varies by Data Science Role}
Perhaps as a testament to the advancement of AutoML technology, participants reported that model building  ``is fast and easy''(P12). P25 further elaborated that 

\begin{quote}
    If I can actually get through all the data stuff, then getting the predictive model is not really that hard. There’s a huge bunch of code to get the data ready for the model, then a tiny bit of code for the model. And then the rest of the work is delivery to the customer.
\end{quote}

The desired amount of oversight and control over AutoML in the analysis appeared to vary by level of technical expertise. P26 felt that individuals with high expertise in statistics and/or computer science were less likely to use automation because ``they want everything to be customizable'', whereas those with less technical expertise, whom they refer to as \textit{``citizen data scientits''} were  ``focused on integrating intelligence to their app'' and tended to prefer higher levels of automation. This latter group is growing as P03 observed : ``the vast majority of data science type work is done by non-data-scientist[s]''. From participants' responses we also noted that this group with a low or evolving technical expertise often needed heavy guidance, low or no-code AutoML implementations, and visualization. \rev{P29 also observed how individuals with higher technical expertise could act as a rate limiting step to analysts engaged in data work, but that automation could serve as a catalyst:}

\begin{quote}
    \rev{[Analysts] maybe can do their own little, their own little predictions, off to the side. And they can do them fast [...] PhDs could always do it better, but are they there? Do they have time? The answer is almost always no. You can either do good or you can do nothing. Better is there, but you're not getting better. They're [PhDs] busy, working on the big problems.}
\end{quote}

\rev{Moreover, summarize that individual with without high technical expertise benefited from proper guidance}, no-code solutions, and data visualization to help situate themselves within the analysis process given many ``steps in (data science) workflow contain lots of details that are hidden'' (P03). 

It was surprising that visualization was not more widely mentioned to steer the model authoring processes even though there exists a number of visualizations systems to help individuals do so~\cite{Chatzimparmpas2020,ChatzimparmpasB_2020,ZhangY2020}. One possibility is that individuals with high technical expertise are constructing novel and bespoke models that existing data visualization tools cannot easily support. Instead, these technical experts might benefit from highly customized visualizations for certain classes of models, such as those generated by tensorboard~\cite{Ham:2018}. In contrast, individuals with lower technical expertise lack the background knowledge to orient themselves and effectively interact with visualizations exposing the mathematical underpinnings of machine learning models. They rely on the AutoML systems to make model decisions and it may not be easy for them to correct and refine these models. 

While participants did raise concerns that AutoML-derived models and results were a `black box', it also appears that technical acumen in statistics and computer science was perceived as necessary and possibly sufficient to `open up the black box'.  Echoing concerns we summarized in Section~\ref{subsec:attidue-bad-decisions} participants saw AutoML as potentially contributing an existing ``knowledge gap [that is] becoming wider'' (P19) because the availability of AutoML meant that individuals with lower technical expertise could harness the power of machine learning without having to seriously engage with its technical and nuanced underpinnings. Our interpretation of these concerns was that trust in the individual conducting the data analysis was as important as trust in the AutoML process itself. Moreover, collaboratively sharing knowledge or including human oversight may mitigate some of these concerns and could be achieved through visualization of data science processes.  \rev{The differences in data science roles and the extension of data science work to individuals without formal training in statistics and computer science has several implications for the design of AutoML and visualization tools. While AutoML tools reduce or even eliminate the need to write code it becomes important to consider what kinds of guard rails might need to be put in place. We believe that data visualization tools are an important component of such guardrails, but that we require a finer-grained understanding of data workers to design such tools effectively.}


\subsubsection{`Good Enough' Rapid Prototyping to Bootstrap Analysis}~\label{subsec:attitude-general}
Participants used AutoML technologies to rapidly prototype viable solutions in both data preparation and analysis. The need for rapid prototyping stems from the challenges of generalizing AutoML to a variety of problems, which requires manual effort. P21 acknowledged that ``AutoML is really hard, and I think we have so many operations with such nuance that we actually most of the time really... just want to be doing simple stuff correctly, rather than adding additional layers of complication.'' P24 was much more explicit in stating that ``every customer is different [...] [but] AutoML is supposed to be a generalized framework. So, that is a problem''. The challenges of AutoML to generalize to a variety of problems are known, especially when it concerns fixed structure computational pipelines (which is the most common implementation of AutoML)~\cite{zoller2019benchmark}. Despite this lack of generalizability, participants found that they could leverage AutoML to rapidly prototype data science solutions. P23 offered description of such a use case: 
\begin{quote}
    I talk to clients daily. If I could get ML done just real quick, as a prototype into what we could build (in depth), that would be super helpful. [...] Can you get me 50\% of the way to answering some question quickly, that would benefit me
\end{quote}

P21 reported using automatically generated results to start a conversation with others ahead of making serious personnel or infrastructure investments. They shared that ``when starting out with a client, we'll run the default model and we'll say, `Hey, here are some of the topics, some of the interesting trends that are coming out' '' and then use the clients reaction to further refine upon default model or craft a different solution altogether. As we previously reported, P12 and P29 also make the case for 'failing-fast' to discover issues in the data or analysis without expensive upfront investment in fully developing an automated data science pipeline.  This rapid and iterative use of AutoML to drive different data conversations evokes a complex picture of a data worker operating within multiple loops of data science and organizational processes. \rev{This prototyping scenario offers an evocative example of how the limitations of AutoML technology can be beneficial leveraged through human-in-the-loop interactions. We argue that with adequate guardrails in place AutoML systems may also be able to further support the process of surfacing potentially more complex issues of bias. }

\subsubsection{Inadequate Support for Governed and Managed Deployment Processes}

The majority of machine learning models often do not advance to production environments where they are applied to real  data. Those that do are often required to go through a set of governed processes before they are deployed and are constantly monitored once they are out in the wild. These governed processes vary as P01 described : 

\begin{quote}
    [A] Governed workflow [is important]. Looking at what all teams are doing - are there divergences around governance, etc. e.g. models with a financial impact have a very stringent governance process.
\end{quote}

The volume and variety of both data and models makes it challenging to monitor and govern AutoML models deployed in production. P03 reported that their practice was to ``err on the side of letting people use tools [they preferred]'' and to  ``monitor what tools are being used''. They emphasized that vigilance was important to ensure mistakes do not ``clog up the server with bad content'', which might happen when pushing a model generated by AutoML into production without adequately vetting it. These problems exist for all software code in general, but may be further exacerbated by the novelty and complexity of AutoML. Moreover, the amount of data produced by automation can make it overwhelming to effectively govern models that need to be continually validated, and have a process that employs someone to ``look for drift, look to increase accuracy and effectiveness of these models over time''(P07). Larger organizations working under enforceable regulatory constraints struggle to find the right balance between integrating a potentially valuable new technology like AutoML while conforming to these constraints: 
\begin{quote}
    [There are] areas where critical models are developed that will likely have very strict controls, often imposed by a regulator [...]if you have no clue what that [model result] means, you are on pretty thin ice (P03)
\end{quote}

As with preparation, the processes of governing deployed models still requires considerable human effort. Moreover, we believe that adding some sort of automation to these processes is desirable in order to reap the efficiency benefits of AutoML. Dashboards are often used to monitor changes in data~\cite{Sarikaya2019}, but participants did not report using dashboards for AutoML work even though they may already use dashboards for other types of work. We hypothesize that this relates to the tooling environment and that monitoring and the governance of AutoML systems require more specialized dashboards that are not well supported by existing tools. There may be fruitful work here for visualization and HCI research to improve governance processes through better monitoring and, at least, help them triage governance violations. \rev{Improved awareness and consideration for governance throughout the visualization design process can also inform the implementation of guardrails for AutoML throughout data science work.}

\subsubsection{Correction and Repair of Deployed Models} Human oversight of automation is critical to detecting when something new or unexpected has happened, identifying the source of what has happened, and implementing appropriate corrective actions as needed. While there may be some ability to automatically detect anomalies, and thus make governed workflows easier to monitor, participants expressed doubt that such an approach would work in practice. P12 stated that if ``the forecast is clearly wrong a human can detect this'' whereas it is harder for the AutoML tool to do so. Still, other participants articulated the limited abilities of analysts to intervene appropriately, suggesting that some analysts ``wouldn't necessarily know what to do next...[whereas]...a data scientist might know what to do next - for improving forecasts, for analyzing how good it might be.''  P26 succinctly summarized that as ``you're only as good as what you debug''. These observations align with a recurrent theme in our findings that trust between the underlying technology and the data science teams is critical for the wider adoption of AutoML. Moreover, these observations expose the brittleness of AutoML technology and its reliance on iterative loops of correction and refinement with humans. In the next section, we also emphasize that AutoML loops are not closed systems; rather these are loops that interact with many other loops of business processes and pre-existing modes of human collaboration. Thus, `debugging' not only requires technical expertise that spans the preparation to deployment processes, but includes sufficient domain expertise to recognize and account for other `loops' that interact with and beyond data science.

\subsubsection{Communication and Collaboration Build Trust in AutoML}\label{subsec:auto_in_ds_collab}

Communication and collaboration are essential data science processes, including but not limited to model automation~\cite{Kim:2017,Wang:2019,zhang:2020}. AutoML systems require communication between humans and the technical system. P22 expresses one such mode of communication in which the AutoML system helps guide users in analysis by communicating ``this is what it is that you're about to do, and this is the impact it will have'' and should also prompt users with respect to certain actions with ``are you sure you want to do this?''. P17 also noted that more could be done to ``walk the user through that [a data analysis, for example] given a kind of data to predict, here are the kinds of models and visualizations to use''. AutoML introduces a new mode of collaboration between humans as well. This new frontier can also be challenging to navigate as P29 observed:
\begin{quote}
So, there's human interaction along the whole life cycle. And interpreting that human interaction is what we're trying to get machine learning to do.
\end{quote}

However, participants indicated that these diverse individuals must still work together to deliver actionable and safe results from automating technology. A common theme emerged around the desire to broaden data engagement across the organization, bringing more people into the data sensemaking loops. In P21's words,
\begin{quote}
    [We] need our workforce to be more data savvy across the board.  An engineer needs to be able to play with data as much as the MBA does [...] [and] giving them better tools will help with ramp up.
\end{quote}

A big part of having teams work more effectively together is to provide more situational awareness of data science workflows and who has done which task. For instance, P06 described what would be needed to support teams working together across workflows. This support includes surfacing "notification[s] that people [are] working on the same step ...[and]... underlying metadata about how people were using the platform '' (P06). Integral to this collaboration was the ability to hand-off different aspects of the data or analysis processes to different team members, likely with a different data science role:
\begin{quote}
They can create a workflow, share it with other people, people can build off of that workflow, grab a table from that workflow and then build their own. That collaboration aspect of it was important to us. (P29)
\end{quote}
In order to improve collaboration some participants defined a viable solution around making workflows visual, interrogable, and extensible.

Participants also highlighted the importance communicating to individuals that were one step removed from the data analysis processes, most often communication to executives or other business leaders

In this process, visualization plays a clear role as a communication tool. A theme emerging from the analysis was that participants often framed this part of the process as more difficult than the modeling itself. This was due in part to the extra work required. For example, P19 describes the additional labor it takes, especially when the visualization tools are not well integrated into existing processes :
\begin{quote}
    There is a gap once your analysis is done on presenting the results.  Nobody wants to spend more hours in another tool to build charts for explanation.
\end{quote}
For instance, P05 described the challenges of authoring compelling visualizations and how ``nice visualizations feel like a hack that the average user can’t build themselves''. 

The other challenge often cited was around the efficacy of visualizations as a medium of communication to drive business decisions or processes. Participants described the challenge of translating their work to business users that were not versed in modeling vernacular. In one participant's words, ``we don't want people to actually understand model jargon, we want to help them understand what the model is saying in business terms.'' This participant relies on interactive visualizations to support the dialogues that they anticipate will happen when showing a snapshot of results to bridge the gap. Still, despite efforts to translate for business users this participant's team experienced a range of challenges in operationalizing their models. P26 describes how ``the challenges that we saw as the data science team is...we give this [model or results] to them, but then actually, the action of implementation of this in the market sometimes doesn't always pull through. So it's like we did all this work, you said it was good, but now you have to take it to the last mile, actually get to marketing, creative, and content, and get it out to market.'' They point to communication difficulties between data scientists and others at the organizations as exacerbating this `last mile' problem, which results from ``either lack of funding or sense of disbelief in prediction models and ML techniques'' (P26).  Validation measures may also be required for regulated industries that can slow this process down.\rev{Taken together, the collaborative nature of data science work imposes constraints on the design of visualization tools, which must be usable across the organization and interoperable by individuals embedded within a variety of analytic, business, and governance processes. }

\vspace{-3mm}
\subsubsection{\rev{Summary}}
\rev{Our examination of AutoML technology along the data science pipeline, both where it exists and where it does not, helps us to understand the current capabilities of this technology and how the technology and its surrounding ecosystem can be further developed to support data scientists and others. We see gaps for AutoML technology outside of data analysis processes and that translate to unmet tooling needs in data preparation, governance, and deployment processes. These are also processes where considerable human labor is still required to make AutoML technology in data analysis viable. Automation that extends to these other processes, ideally with appropriate guardrails, could improve both the quality and speed of data science work. Moreover, the relationship between automation and data science 
expertise emerged as a critical consideration for what future tools should support, including the types of guardrails that should be built in. We were surprised to surface some of the tensions that existed between data science experts,
and data workers with different training. Emerging from this tension was one heavy-handed guardrail strategy to restrict access to AutoML technology that many sought to implement. We believe this view has surfaced from a lack of adequate tools to support the safe creation, deployment, and governance of these models and that there are many fertile opportunities for visualization research in this space. However, our analysis of participants' comments reveals that existing visualization tools are falling short of their needs. Moreover, that data visualizations tools can have a steep learning curve and their is little motivation to use following intensive analysis. We underscore that it is critical to understand the diversity of teams that carry out data science work and the ways they intersect with many organizational processes. In other words, visualization tools need to work for many humans engaged in many loops.}
\section{Interpretation of Findings}
We now reflect on our findings and summarize the central themes that emerged from our analysis.

\subsection{Three  Usage Scenarios for AutoML Emerge }\label{subsec:attitude-summary}
The general attitudes toward AutoML suggest three usage scenarios for this technology that are conditioned on the technical expertise (statistics and computer science) of the individual analyzing the data and the magnitude of consequences associated with errors. The first usage scenario is automating routine tasks, thereby reducing the coding efforts of data science teams and improving the speed of the analysis processes. A second usage scenario is the rapid exploration of potential data science solutions through low-effort prototyping. Such prototyping approaches can be used by individuals with varying degrees of technical expertise. \rev{Its possible that for individuals with high technical expertise (such as, data scientists, generalists, research scientists, ML/AI engineers, and data shapers)  prototyping allows them to quickly create a base framework that they further develop into novel solutions for arising technical challenges. For other individuals, prototyping enables them to have a conversation around the data with customers and other members of their organization. Prototyping also enables individuals and data science teams to fail fast and discover issues with their data and analysis before investing in considerable engineering effort.} A third and final usage scenario is the use of AutoML toward democratizing the ability to create a machine learning model, empowering individuals that would not be able to build a model otherwise.  In this third scenario, we argue that individuals require heavy guidance  and guardrails from an AutoML systems and may have very limited ability to identify errors or correct them. 

The delineation of these usage scenarios is intended to guide visualization researchers as they explore opportunities to develop techniques or systems for AutoML. 

\subsection{Varying the Level of Automation Across Data Science Processes}
Considerable human labor is still expended to prepare data, govern and deploy a model, and to communicate the results to impacted individuals and other decision-makers. An end-to-end AutoML solution capable of addressing the full scope of such data science work does not currently exist, and as a result data workers, which includes individuals that are and are not data scientists, are finding \textit{ad hoc} ways of bootstrapping AutoML technology into their work.  In~\autoref{fig:info_seek}, we outline a common set of eight steps synthesized from participants' responses describing AutoML use in enterprise settings. We further align these steps within higher order data science processes. For data preparation and analysis, these tasks were prototyping, exploring the results, and settling on a solution to implement. Should this solution reach a certain level of maturity it is deployed into production following a verification of the solution (including compliance of regulations), where it is consistently monitored while in production. Finally, these deployed models can be used to take action through an inspection of the results that surface new insights for decision making.  We illustrate the levels of automation~\cite{Parasuraman2000} that we believe are desirable for future AutoML systems to support, considering the range of participant challenges and concerns this study surfaced. Importantly, the level of automation is not consistent across all data science processes. Human oversight is still required throughout data science work and is dictated by both regulatory requirements and organizational practices.  Most automation likely needs to adopt a `cruise control' mode of interaction~\cite{Lee2019AHP}, where humans can oversee and steer AutoML systems without needing to guide the systems at each step. Even this would be an improvement over current AutoML systems that appear to oscillate between 'autopilot' and 'user-driven' modes. 
We further illustrate the level of automation required by individuals with high expertise in computer science and/or statistics (Data Scientists, ML/AI  or Data Engineers), and low or an evolving technical expertise in these areas (Business Analysts, Moonlighters). Individuals with high expertise can benefit from full automation, for example when speeding up routine work (Usage Scenario One) or to rapidly prototype and explore new solutions (Usage Scenario Two). 
Even in these two usage scenarios individuals with high technical expertise still rely on considerable manual effort, but this in fact might be an appropriate use of their expertise and focus on ``bigger problems'', especially if other trivially automated tasks are reliably handled by an AutoML system. Individuals with lower or and evolving technical expertise require much more support and guidance and would rely much more on full automation to rapidly prototype solutions (Usage Scenario Two) or even to begin to engage in data science work more generally (Usage Scenario Three). However, while these individuals rely on AutoML systems to guide them, their domain expertise still needs to be incorporated in downstream steps. 

While~\autoref{fig:info_seek} is a useful illustrative summary of our findings, it needs to be further validated in future studies that assess its generalizability. We suggest how to do so in our Discussion section. 

\subsection{\rev{Eliciting Tasks for Visualization Design}}
\rev{Taken together, these usage scenarios an levels of automation impose a set of constraints for the design of visualization tools that operate together with AutoML technologies. Visualization researchers need to carefully consider where and how automation is currently deployed, the diversity and expertise of the data science teams, and the full breadth
of data science processes. We have illustrated a set of steps and proposed the levels of automation in~\autoref{fig:info_seek} that data workers with different levels of expertise desire. Importantly, by illustrating an end-to-end pipeline, we encourage visualization researchers to consider how changes \textit{across a workflow} influence the kinds of data to be visualized and the fundamental tasks that these workflow steps support. For example, `prototyping' may have different tasks associated to it depending on whether the analysts want to develop a new model, fail fast, or prototype some solution for a customer. The `monitor' process in deployment could reasonably rely on high automation until the system requires human action, much like auto pilot in aircraft. Alternatively, `exploration' may require less automation if the user is expected to steer the algorithm. Without a concrete understanding of usage scenarios, data science steps, and level of automation, researchers risk eliciting inappropriate tasks and creating visualization tools that will be dismissed because they are not well integrated into end-to-end data science workflows.  Visualization researchers can reference our findings and the summary in~\autoref{fig:info_seek} as a guide to support their own task elicitation for the design and evaluation of visualization tools.}



\subsection{Modifications to our Analytic Framework}\label{sec:framework-modifications}
Lastly, we briefly reflect on our findings and propose modifications to the framework of data science work and workers reported in~\cite{Crisan_2021_dsframework}. We remind the reader that this framework is described in Section~\ref{subsec:methods-analysis} and delineates a set of \textbf{higher} and associated \textit{lower} order data science processes that we used as part of selective coding analysis. First, we propose that \textit{Collaboration} be added as a lower-order processes of \textbf{Communication}. While collaboration was part of the original framework there was not enough evidence to determine how it should be incorporated. This analysis suggests it belongs as a component of \textbf{Communication} alongside \textit{documentation} and \textit{dissemination} lower order processes. Moreover, \textit{collaboration} emphasizes the ways that individuals engage in multi-directional exchanges of knowledge and data products (data, code, models, documents), whereas \textit{dissemination} refers to a more unidirectional exchange of knowledge from an individual to others. Second, we propose that \textit{governance} be included as a lower-order process of \textit{deployment}. While governance processes can technically encompass all of data science work, our findings point to its specific importance in managing the process of launching, monitoring, and refining machine learning models deployed into production settings. Finally, we propose a new higher order process, \textbf{Guidance}, which follows communication. We assign three lower order guidance processes based upon our analysis : \textit{human-machine guidance}, \textit{human-human guidance} (or pedagogy), and \textit{organizational guidance}. \textit{Human-machine guidance} describes the interplay between AutoML tools surfacing new data insights to humans and humans making corrections and refinements of AutoML models and results. \textit{Human-human} guidance describes the collective work in building a data savvy organization and other efforts to bridge the data science ``knowledge gap''. Alternatively this could be referred to as pedagogical process. Finally, \textit{organization guidance} refers to regulations and other organizational processes that impose constraints on the use of data, models, and the level of automation.

\section{Discussion}\label{sec:discussion}
Visualization and HCI researchers have used enterprise studies to discover unmet needs of practitioners that have inspired new research trajectories that have ultimately led to new techniques and tools.  As we consider the future of AutoML in enterprise, we believe a ``cruise control'' level of interaction~\cite{Lee2019AHP} (\autoref{fig:info_seek}) is more likely to be adopted. However, we see significant barriers to implementing such a level of automation that stem from the diversity among data workers with different types expertise, a complex tooling environment that needs to be integrated, and brittle workflows that still rely on considerable human effort. Although visualization can play a role in supporting `cruise-control' type automation, it was not being widely used to that effect and, in some cases, getting actively removed from automated data science workflows.\rev{We believe this lack of uptake is that visualization tools are potentially misspecified for the tasks they need to support and that this stems from poor understand of how automation is used in data science work and where there are opportunities for human-in-the-loop interaction. Our study fills this gap by surfacing usage scenarios and illustration of automation throughout data science work, which informs the goals and tasks feeding into visualization design and evaluation.}

\subsection{Implications for Automating Data Science Work}
Throughout our analysis, we found both AutoML and human-in-the-loop to be misnomers for the processes that participants were describing. First, we noted in Section~\ref{sec:related_work} that AutoML is used to refer to an ever-expanding set of data science processes from preparation to deployment and as such is being used interchangeably with `automating data science' (among other phrases). We argue this is limiting as not all automation of data science needs to be in service of machine learning systems. Moreover, the notion of end-to-end AutoML obscures the human labor required for these systems to work, now and in the future, leaving inadequate support for human-machine collaboration. Echoing Wang's~\cite{Wang:2019} language, we encourage researchers to \textbf{augmenting data science} with AutoML rather than automating it. It is more than a matter of semantics -- the idea of augmenting data work explicitly makes space for human engagement and brings humans needs to the forefront of consideration. 

Second, when we make explicit space for human engagement we are encouraged to consider the diversity among data workers. As we summarize in our three usage scenarios, this type of engagement will vary depend on the goals of data workers and their level of technical expertise. Along with prior studies~\cite{zhang:2020,Passi2018} we found collaboration among data workers to be of critical importance to the success of data work. Commensurate with findings from Hong~\cite{Hong_2020} we also show that \textbf{trust amongst individuals engaged in data work was as important, or more so, than trust in AutoML.} Surprisingly, AutoML technology appeared to erode trust among collaborators of different technical expertise by enabling so called ``citizen data scientists'' to potentially automate bad decision making. In theory, a human-in-the-loop paradigm for augmenting data science work can also be useful to understand the types of engagement between humans and machines that could ameliorate some of these trust concerns. However, here, too we find that human-in-the-loop is a limiting term. An AutoML correction and refinement loop not only exists within a wider scope of data science processes but also within organizational processes. While the nomenclature of human-in-the-loop is not exclusive to a single individual interacting with AutoML, we argue that the notion of ``humans-in-the-loops'' more accurately captures how this technology is used within enterprise settings. \rev{We note that a limitation of our findings was that study participants were primarily, although with some exceptions, experts in data science. 
While several were managers who oversaw mixed teams, we none-the-less believe it is useful to follow-up our findings by soliciting the views of those individuals that are not data scientists, but work closely with them}.

As Visualization and HCI researchers continue to explore applications of technology like AutoML in data science work, we encourage them to consider the diversity of humans involved in data science work, their different needs and varying degrees to which they benefit from AutoML technology as well as the myriad organizational loops that are entangled within AutoML and data science. 

\subsection{Implications for Data Visualization Systems}
Overall, we see that there are opportunities for visualization tools in data science work, especially in areas where there already exist considerable human labor. We especially see that participants struggle to get an overview of data work and that this complicates their ability to effectively 
handoff data, models, and results within their organizations. A visual overview of data science workflows emerged as an organic solution and is a promising area of future research. But beyond this specific example, we hope that the usage scenarios we present will help researchers identify new unmet visualization needs toward the use of AutoML that we did not surface here. However, the most troubling findings from our study concern the ecological validity of data visualization systems. We hypothesize that one reason visualization tools were not more widely used by participants was because they did not integrate well into existing data science tooling environments. This may be because existing visualization tools are developed as stand-alone systems where it is difficult to import data and export results, or because existing systems do not scale well to the volumes and varieties of data that organizations collect, or even because these visualization systems are themselves too brittle to flexibly adapt to variable data science or AutoML workflows. Moreover, visualization tools may not cater well to individuals across the gradient of technical expertise, and thus may be too rudimentary for those with high technical expertise and too complex for those with lower expertise.  We encourage researchers to use our findings as a guide for surfacing these threats of ecological validity early. 

\rev{Another fruitful area for visualization researchers is the creation of guardrails that surface and alert individuals of potential issues with their data, models, or results. The development of guardrails can help to examine concerns toward automating bad decisions. Our research indicates that their design is contingent upon individual expertise, the context in which individuals are using AutoML, and the level of automation that is expected. Some areas, like data preparation, will require more human labor alongside tools that automate their processes. Others, like monitoring a deployment model would rely on human labor primarily to respond to events, like the detection of model drift. Guardrails in both scenarios can help analysts contextualize and triage problems as they arise, but the design of these guardrails will differ between these two scenarios. Well designed guardrails may also increase trust and collaboration not only between data workers and automated processes, but also among data science teams. While prior research has suggested design considerations~\cite{Amershi2019} and potential analytic pitfalls across visual analytics processes~\cite{McNutt2020}, research is needed to bring these together to explore dynamic and adaptive visualization guardrails that are appropriate for an individual's current analytic context. }

\subsection{Limitations and Future Work}
The lack of existing studies on AutoML use in enterprise settings was the motivating factor for carrying out this research. Our findings support prior research and shed new light on the challenges and uses of AutoML in enterprise settings. However, we also found that participants had quite different experiences in their use and expectations of AutoML. As a result, our findings were simultaneously rich in capturing the diversity of experiences and sparse in that some of our findings relied on a handful of observations. To produce a cohesive analysis of these experiences we used an existing framework for data science work and workers as a scaffold. This sparseness of data and reliance on a scaffold is the primary limitation of our findings. Further work is needed to validate the generalizability of our findings, but this may be difficult due to the novelty of AutoML technology itself. One fruitful area of future work is to take the key insights from our research as constructs around which to develop a survey instrument that probes into AutoML uses more specifically than our current interview study. We did not take this approach here because we felt we needed additional information on AutoML use in the enterprise settings and beyond. A future survey instrument could also be used within a large mixed-methods approach, such as sequential explanatory design, which uses the survey results, in lieu of the framework we use here, as a more data-driven approach to inform a subsequent qualitative analysis.

\section{Conclusion}
Automating data science work through AutoML technology will continue to be commonplace in enterprise settings, especially at large organizations that work with large volumes of data. We identified three usage scenarios for AutoML that we argue are routine in current enterprise environments. These are automation routine work, rapid prototyping for a potential solution, and democratizing access to machine learning technology and data science work more generally. Moreover, we surface the complex handoff of data work between AutoML systems and data workers, as well as between data workers having different levels of technical expertise. Indeed, AutoML systems still rely on considerable human effort to be effective and even as this technology improves, human oversight will still be required to be sure it is safe and effective. While data visualization can play an important role together with AutoML, we find that it is used infrequently and is actively being minimized in data science work. We see our findings as having important implications for recasting the role of visualization in conjunction with AutoML and data science more generally. 

\begin{acks}
The authors wish to acknowledge and thank to study participants for sharing their insights with us. We also wish to acknowledge members of the Tableau Research, User Research, and Tableau CRM for their feedback on our study and findings. 
\end{acks}

\bibliographystyle{ACM-Reference-Format}
\bibliography{main}


\begin{thebibliography}{54}


\ifx \showCODEN    \undefined \def \showCODEN     #1{\unskip}     \fi
\ifx \showDOI      \undefined \def \showDOI       #1{#1}\fi
\ifx \showISBNx    \undefined \def \showISBNx     #1{\unskip}     \fi
\ifx \showISBNxiii \undefined \def \showISBNxiii  #1{\unskip}     \fi
\ifx \showISSN     \undefined \def \showISSN      #1{\unskip}     \fi
\ifx \showLCCN     \undefined \def \showLCCN      #1{\unskip}     \fi
\ifx \shownote     \undefined \def \shownote      #1{#1}          \fi
\ifx \showarticletitle \undefined \def \showarticletitle #1{#1}   \fi
\ifx \showURL      \undefined \def \showURL       {\relax}        \fi
\providecommand\bibfield[2]{#2}
\providecommand\bibinfo[2]{#2}
\providecommand\natexlab[1]{#1}
\providecommand\showeprint[2][]{arXiv:#2}

\bibitem[\protect\citeauthoryear{Alspaugh, Zokaei, Liu, Jin, and
  Hearst}{Alspaugh et~al\mbox{.}}{2019}]%
        {Alspaug:2018}
\bibfield{author}{\bibinfo{person}{Sarah Alspaugh}, \bibinfo{person}{Nava
  Zokaei}, \bibinfo{person}{Andrew Liu}, \bibinfo{person}{Cindy Jin}, {and}
  \bibinfo{person}{Marti~A. Hearst}.} \bibinfo{year}{2019}\natexlab{}.
\newblock \showarticletitle{Futzing and Moseying: Interviews with Professional
  Data Analysts on Exploration Practices}.
\newblock \bibinfo{journal}{\emph{IEEE Transactions on Visualization and
  Computer Graphics}} \bibinfo{volume}{25}, \bibinfo{number}{1}
  (\bibinfo{year}{2019}), \bibinfo{pages}{22--31}.
\newblock
\urldef\tempurl%
\url{https://doi.org/10.1109/TVCG.2018.2865040}
\showDOI{\tempurl}


\bibitem[\protect\citeauthoryear{Amershi, Weld, Vorvoreanu, Fourney, Nushi,
  Collisson, Suh, Iqbal, Bennett, Inkpen, Teevan, Kikin-Gil, and
  Horvitz}{Amershi et~al\mbox{.}}{2019}]%
        {Amershi2019}
\bibfield{author}{\bibinfo{person}{Saleema Amershi}, \bibinfo{person}{Dan
  Weld}, \bibinfo{person}{Mihaela Vorvoreanu}, \bibinfo{person}{Adam Fourney},
  \bibinfo{person}{Besmira Nushi}, \bibinfo{person}{Penny Collisson},
  \bibinfo{person}{Jina Suh}, \bibinfo{person}{Shamsi Iqbal},
  \bibinfo{person}{Paul~N. Bennett}, \bibinfo{person}{Kori Inkpen},
  \bibinfo{person}{Jaime Teevan}, \bibinfo{person}{Ruth Kikin-Gil}, {and}
  \bibinfo{person}{Eric Horvitz}.} \bibinfo{year}{2019}\natexlab{}.
\newblock \showarticletitle{Guidelines for Human-AI Interaction}. In
  \bibinfo{booktitle}{\emph{Proc CHI'19}}. \bibinfo{pages}{1–13}.
\newblock
\urldef\tempurl%
\url{https://doi.org/10.1145/3290605.3300233}
\showDOI{\tempurl}


\bibitem[\protect\citeauthoryear{Blei and Smyth}{Blei and Smyth}{2017}]%
        {Blei2017}
\bibfield{author}{\bibinfo{person}{David~M. Blei} {and}
  \bibinfo{person}{Padhraic Smyth}.} \bibinfo{year}{2017}\natexlab{}.
\newblock \showarticletitle{Science and Data Science}.
\newblock \bibinfo{journal}{\emph{Proceedings of the National Academy of
  Sciences}} \bibinfo{volume}{114}, \bibinfo{number}{33}
  (\bibinfo{year}{2017}), \bibinfo{pages}{8689--8692}.
\newblock
\urldef\tempurl%
\url{https://doi.org/10.1073/pnas.1702076114}
\showDOI{\tempurl}


\bibitem[\protect\citeauthoryear{Bowen}{Bowen}{2006}]%
        {Bowen}
\bibfield{author}{\bibinfo{person}{Glenn~A. Bowen}.}
  \bibinfo{year}{2006}\natexlab{}.
\newblock \showarticletitle{Grounded Theory and Sensitizing Concepts}.
\newblock \bibinfo{journal}{\emph{International Journal of Qualitative
  Methods}} \bibinfo{volume}{5}, \bibinfo{number}{3} (\bibinfo{year}{2006}),
  \bibinfo{pages}{12--23}.
\newblock
\urldef\tempurl%
\url{https://doi.org/10.1177/160940690600500304}
\showDOI{\tempurl}


\bibitem[\protect\citeauthoryear{Bryant and Charmaz}{Bryant and
  Charmaz}{2007}]%
        {Bryant_Charmaz_2011}
\bibfield{author}{\bibinfo{person}{Anthony Bryant} {and} \bibinfo{person}{Kathy
  Charmaz}.} \bibinfo{year}{2007}\natexlab{}.
\newblock \bibinfo{booktitle}{\emph{The SAGE Handbook of Grounded Theory}}.
\newblock \bibinfo{publisher}{Sage Publications}, \bibinfo{address}{Los
  Angeles, Calif.}
\newblock
\showISBNx{978-1-4129-2346-0}


\bibitem[\protect\citeauthoryear{Cao}{Cao}{2017}]%
        {Longbing:2017}
\bibfield{author}{\bibinfo{person}{Longbing Cao}.}
  \bibinfo{year}{2017}\natexlab{}.
\newblock \showarticletitle{Data Science: A Comprehensive Overview}.
\newblock \bibinfo{journal}{\emph{Comput. Surveys}} \bibinfo{volume}{50},
  \bibinfo{number}{3} (\bibinfo{year}{2017}), \bibinfo{pages}{1--42}.
\newblock
\urldef\tempurl%
\url{https://doi.org/10.1145/3076253}
\showDOI{\tempurl}


\bibitem[\protect\citeauthoryear{Chatzimparmpas, Martins, Jusufi, and
  Kerren}{Chatzimparmpas et~al\mbox{.}}{2020a}]%
        {Chatzimparmpas2020}
\bibfield{author}{\bibinfo{person}{Angelos Chatzimparmpas},
  \bibinfo{person}{Rafael~M. Martins}, \bibinfo{person}{Ilir Jusufi}, {and}
  \bibinfo{person}{Andreas Kerren}.} \bibinfo{year}{2020}\natexlab{a}.
\newblock \showarticletitle{A Survey of Surveys on the Use of Visualization for
  Interpreting Machine Learning Models}.
\newblock \bibinfo{journal}{\emph{Information Visualization}}
  \bibinfo{volume}{19}, \bibinfo{number}{3} (\bibinfo{year}{2020}),
  \bibinfo{pages}{207--233}.
\newblock
\urldef\tempurl%
\url{https://doi.org/10.1177/1473871620904671}
\showDOI{\tempurl}


\bibitem[\protect\citeauthoryear{Chatzimparmpas, Martins, Jusufi, Kostiantyn,
  Fabrice, and Kerren}{Chatzimparmpas et~al\mbox{.}}{2020b}]%
        {ChatzimparmpasB_2020}
\bibfield{author}{\bibinfo{person}{Angelos Chatzimparmpas},
  \bibinfo{person}{Rafael~M. Martins}, \bibinfo{person}{Ilir Jusufi},
  \bibinfo{person}{Kucher Kostiantyn}, \bibinfo{person}{Rossi Fabrice}, {and}
  \bibinfo{person}{Andreas Kerren}.} \bibinfo{year}{2020}\natexlab{b}.
\newblock \showarticletitle{The State of the Art in Enhancing Trust in Machine
  Learning Models with the Use of Visualizations}.
\newblock \bibinfo{journal}{\emph{Computer Graphics Forum}}
  \bibinfo{volume}{39}, \bibinfo{number}{3} (\bibinfo{year}{2020}),
  \bibinfo{pages}{713–756}.
\newblock
\urldef\tempurl%
\url{https://doi.org/10.1111/cgf.14034}
\showDOI{\tempurl}


\bibitem[\protect\citeauthoryear{Creswell and Poth}{Creswell and Poth}{2018}]%
        {Creswell_Poth_2018}
\bibfield{author}{\bibinfo{person}{John~W. Creswell} {and}
  \bibinfo{person}{Cheryl~N. Poth}.} \bibinfo{year}{2018}\natexlab{}.
\newblock \bibinfo{booktitle}{\emph{Qualitative inquiry \& Research Design:
  Choosing Among Five Approaches} (\bibinfo{edition}{fourth edition} ed.)}.
\newblock \bibinfo{publisher}{Sage Publications}, \bibinfo{address}{Los
  Angeles, Calif}.
\newblock
\showISBNx{978-1-5063-3020-4}


\bibitem[\protect\citeauthoryear{Crisan, Fiore-Gartland, and Tory}{Crisan
  et~al\mbox{.}}{2020}]%
        {Crisan_2021_dsframework}
\bibfield{author}{\bibinfo{person}{Anamaria Crisan}, \bibinfo{person}{Brittany
  Fiore-Gartland}, {and} \bibinfo{person}{Melanie Tory}.}
  \bibinfo{year}{2020}\natexlab{}.
\newblock \showarticletitle{Passing the Data Baton: A Retrospective Analysis on
  Data Science Work and Workers}.
\newblock \bibinfo{journal}{\emph{IEEE Transactions on Visualization and
  Computer Graphics}} (\bibinfo{year}{2020}).
\newblock
\urldef\tempurl%
\url{https://doi.org/10.1109/TVCG.2020.3030340}
\showDOI{\tempurl}


\bibitem[\protect\citeauthoryear{Crisan and Munzner}{Crisan and
  Munzner}{2019}]%
        {2020_datarecon}
\bibfield{author}{\bibinfo{person}{Anamaria Crisan} {and}
  \bibinfo{person}{Tamara Munzner}.} \bibinfo{year}{2019}\natexlab{}.
\newblock \showarticletitle{Uncovering Data Landscapes through Data
  Reconnaissance and Task Wrangling}.
\newblock \bibinfo{journal}{\emph{2019 IEEE Visualization Conference (VIS)}},
  \bibinfo{pages}{46--50}.
\newblock
\urldef\tempurl%
\url{https://doi.org/10.1109/VISUAL.2019.8933542}
\showDOI{\tempurl}


\bibitem[\protect\citeauthoryear{Donoho}{Donoho}{2017}]%
        {Donoho:2017}
\bibfield{author}{\bibinfo{person}{David Donoho}.}
  \bibinfo{year}{2017}\natexlab{}.
\newblock \showarticletitle{50 Years of Data Science}.
\newblock \bibinfo{journal}{\emph{Journal of Computational and Graphical
  Statistics}} \bibinfo{volume}{26}, \bibinfo{number}{4}
  (\bibinfo{year}{2017}), \bibinfo{pages}{745--766}.
\newblock
\urldef\tempurl%
\url{https://doi.org/10.1080/10618600.2017.1384734}
\showDOI{\tempurl}


\bibitem[\protect\citeauthoryear{Drozdal, Weisz, Wang, Dass, Yao, Zhao, Muller,
  Ju, and Su}{Drozdal et~al\mbox{.}}{2020}]%
        {Drozal2020}
\bibfield{author}{\bibinfo{person}{Jaimie Drozdal}, \bibinfo{person}{Justin
  Weisz}, \bibinfo{person}{Dakuo Wang}, \bibinfo{person}{Gaurav Dass},
  \bibinfo{person}{Bingsheng Yao}, \bibinfo{person}{Changruo Zhao},
  \bibinfo{person}{Michael Muller}, \bibinfo{person}{Lin Ju}, {and}
  \bibinfo{person}{Hui Su}.} \bibinfo{year}{2020}\natexlab{}.
\newblock \showarticletitle{Trust in AutoML: Exploring Information Needs for
  Establishing Trust in Automated Machine Learning Systems}. In
  \bibinfo{booktitle}{\emph{Proc IUI'20}}. \bibinfo{pages}{297–307}.
\newblock
\urldef\tempurl%
\url{https://doi.org/10.1145/3377325.3377501}
\showDOI{\tempurl}


\bibitem[\protect\citeauthoryear{Fern\'{a}ndez-Delgado, Cernadas, Barro, and
  Amorim}{Fern\'{a}ndez-Delgado et~al\mbox{.}}{2014}]%
        {Manuel2014}
\bibfield{author}{\bibinfo{person}{Manuel Fern\'{a}ndez-Delgado},
  \bibinfo{person}{Eva Cernadas}, \bibinfo{person}{Sen\'{e}n Barro}, {and}
  \bibinfo{person}{Dinani Amorim}.} \bibinfo{year}{2014}\natexlab{}.
\newblock \showarticletitle{Do we Need Hundreds of Classifiers to Solve Real
  World Classification Problems?}
\newblock \bibinfo{journal}{\emph{Journal of Machine Learning Research}}
  \bibinfo{volume}{15}, \bibinfo{number}{90} (\bibinfo{year}{2014}),
  \bibinfo{pages}{3133--3181}.
\newblock
\urldef\tempurl%
\url{http://jmlr.org/papers/v15/delgado14a.html}
\showURL{%
\tempurl}


\bibitem[\protect\citeauthoryear{Feurer, Eggensperger, Falkner, Lindauer, and
  Hutter}{Feurer et~al\mbox{.}}{2020}]%
        {autosklearn2020}
\bibfield{author}{\bibinfo{person}{Matthias Feurer}, \bibinfo{person}{Katharina
  Eggensperger}, \bibinfo{person}{Stefan Falkner}, \bibinfo{person}{Marius
  Lindauer}, {and} \bibinfo{person}{Frank Hutter}.}
  \bibinfo{year}{2020}\natexlab{}.
\newblock \bibinfo{title}{Auto-Sklearn 2.0: The Next Generation}.
\newblock
\newblock
\urldef\tempurl%
\url{https://arxiv.org/abs/2007.04074}
\showURL{%
\tempurl}


\bibitem[\protect\citeauthoryear{Feurer, Klein, Eggensperger, Springenberg,
  Blum, and Hutter}{Feurer et~al\mbox{.}}{2015}]%
        {autosklearn2015}
\bibfield{author}{\bibinfo{person}{Matthias Feurer}, \bibinfo{person}{Aaron
  Klein}, \bibinfo{person}{Katharina Eggensperger},
  \bibinfo{person}{Jost~Tobias Springenberg}, \bibinfo{person}{Manuel Blum},
  {and} \bibinfo{person}{Frank Hutter}.} \bibinfo{year}{2015}\natexlab{}.
\newblock \showarticletitle{Efficient and Robust Automated Machine Learning}.
  In \bibinfo{booktitle}{\emph{Proc NeurIPs'15}}. \bibinfo{pages}{2755–2763}.
\newblock
\urldef\tempurl%
\url{https://doi.org/10.5555/2969442.2969547}
\showDOI{\tempurl}


\bibitem[\protect\citeauthoryear{Gil, Honaker, Gupta, Ma, D'Orazio, Garijo,
  Gadewar, Yang, and Jahanshad}{Gil et~al\mbox{.}}{2019}]%
        {Gil2019}
\bibfield{author}{\bibinfo{person}{Yolanda Gil}, \bibinfo{person}{James
  Honaker}, \bibinfo{person}{Shikhar Gupta}, \bibinfo{person}{Yibo Ma},
  \bibinfo{person}{Vito D'Orazio}, \bibinfo{person}{Daniel Garijo},
  \bibinfo{person}{Shruti Gadewar}, \bibinfo{person}{Qifan Yang}, {and}
  \bibinfo{person}{Neda Jahanshad}.} \bibinfo{year}{2019}\natexlab{}.
\newblock \showarticletitle{Towards Human-Guided Machine Learning}. In
  \bibinfo{booktitle}{\emph{Proc IUI'19}}. \bibinfo{pages}{614–624}.
\newblock
\urldef\tempurl%
\url{https://doi.org/10.1145/3301275.3302324}
\showDOI{\tempurl}


\bibitem[\protect\citeauthoryear{Golovin, Solnik, Moitra, Kochanski, Karro, and
  Sculley}{Golovin et~al\mbox{.}}{2017}]%
        {googleVizer2017}
\bibfield{author}{\bibinfo{person}{Daniel Golovin}, \bibinfo{person}{Benjamin
  Solnik}, \bibinfo{person}{Subhodeep Moitra}, \bibinfo{person}{Greg
  Kochanski}, \bibinfo{person}{John Karro}, {and} \bibinfo{person}{D.
  Sculley}.} \bibinfo{year}{2017}\natexlab{}.
\newblock \showarticletitle{Google Vizier: A Service for Black-Box
  Optimization}. In \bibinfo{booktitle}{\emph{Proc KDD'17}}.
  \bibinfo{pages}{1487–1495}.
\newblock
\urldef\tempurl%
\url{https://doi.org/10.1145/3097983.3098043}
\showDOI{\tempurl}


\bibitem[\protect\citeauthoryear{Gray and Suri}{Gray and Suri}{2019}]%
        {Gray_Suri_2019}
\bibfield{author}{\bibinfo{person}{Mary~L. Gray} {and}
  \bibinfo{person}{Siddharth Suri}.} \bibinfo{year}{2019}\natexlab{}.
\newblock \bibinfo{booktitle}{\emph{Ghost Work: How to Stop Silicon Valley from
  Building a New Global Underclass}}.
\newblock \bibinfo{publisher}{Houghton Mifflin Harcourt}.
\newblock
\showISBNx{978-1-328-56624-9}


\bibitem[\protect\citeauthoryear{Heer}{Heer}{2019}]%
        {Heer_2019}
\bibfield{author}{\bibinfo{person}{Jeffrey Heer}.}
  \bibinfo{year}{2019}\natexlab{}.
\newblock \showarticletitle{Agency Plus Automation: Designing Artificial
  Intelligence into Interactive Systems}.
\newblock \bibinfo{journal}{\emph{Proceedings of the National Academy of
  Sciences}} \bibinfo{volume}{116}, \bibinfo{number}{6} (\bibinfo{year}{2019}),
  \bibinfo{pages}{1844–1850}.
\newblock
\urldef\tempurl%
\url{https://doi.org/10.1073/pnas.1807184115}
\showDOI{\tempurl}


\bibitem[\protect\citeauthoryear{Honeycutt, Nourani, and Ragan}{Honeycutt
  et~al\mbox{.}}{2020}]%
        {honeycutt2020}
\bibfield{author}{\bibinfo{person}{Donald Honeycutt}, \bibinfo{person}{Mahsan
  Nourani}, {and} \bibinfo{person}{Eric Ragan}.}
  \bibinfo{year}{2020}\natexlab{}.
\newblock \showarticletitle{Soliciting Human-in-the-Loop User Feedback for
  Interactive Machine Learning Reduces User Trust and Impressions of Model
  Accuracy}.
\newblock \bibinfo{journal}{\emph{Proc AAAI HCOMP'2020}} \bibinfo{volume}{8},
  \bibinfo{number}{1}, \bibinfo{pages}{63--72}.
\newblock
\urldef\tempurl%
\url{https://ojs.aaai.org/index.php/HCOMP/article/view/7464}
\showURL{%
\tempurl}


\bibitem[\protect\citeauthoryear{Hong, Hullman, and Bertini}{Hong
  et~al\mbox{.}}{2020}]%
        {Hong_2020}
\bibfield{author}{\bibinfo{person}{Sungsoo~Ray Hong}, \bibinfo{person}{Jessica
  Hullman}, {and} \bibinfo{person}{Enrico Bertini}.}
  \bibinfo{year}{2020}\natexlab{}.
\newblock \showarticletitle{Human Factors in Model Interpretability: Industry
  Practices, Challenges, and Needs}.
\newblock \bibinfo{journal}{\emph{Proc CSCW'2020}}, Article
  \bibinfo{articleno}{068} (\bibinfo{year}{2020}),
  \bibinfo{numpages}{26}~pages.
\newblock
\urldef\tempurl%
\url{https://doi.org/10.1145/3392878}
\showDOI{\tempurl}


\bibitem[\protect\citeauthoryear{Horvitz}{Horvitz}{1999}]%
        {Horvitz_1999}
\bibfield{author}{\bibinfo{person}{Eric Horvitz}.}
  \bibinfo{year}{1999}\natexlab{}.
\newblock \showarticletitle{Principles of Mixed-Initiative User Interfaces}. In
  \bibinfo{booktitle}{\emph{Proc CHI'99}}. \bibinfo{pages}{159–166}.
\newblock
\urldef\tempurl%
\url{https://www.microsoft.com/en-us/research/publication/principles-mixed-initiative-user-interfaces-2/}
\showURL{%
\tempurl}


\bibitem[\protect\citeauthoryear{Inc.}{Inc.}{2020a}]%
        {sagemaker2020}
\bibfield{author}{\bibinfo{person}{Amazon Inc.}}
  \bibinfo{year}{2020}\natexlab{a}.
\newblock \bibinfo{title}{{Amazon SageMaker Autopilot}}.
\newblock
  \bibinfo{howpublished}{\url{https://aws.amazon.com/sagemaker/autopilot/}}.
\newblock
\newblock
\shownote{Accessed: 2020-09-01.}


\bibitem[\protect\citeauthoryear{Inc.}{Inc.}{2020b}]%
        {datarobot2020}
\bibfield{author}{\bibinfo{person}{DataRobot Inc.}}
  \bibinfo{year}{2020}\natexlab{b}.
\newblock \bibinfo{title}{{DataRobot: Empowering the Human Heroes of the
  Intelligence Revolution}}.
\newblock \bibinfo{howpublished}{\url{https://www.datarobot.com/}}.
\newblock
\newblock
\shownote{Accessed: 2020-09-01.}


\bibitem[\protect\citeauthoryear{Inc.}{Inc.}{2020c}]%
        {googleAutoML2020}
\bibfield{author}{\bibinfo{person}{Google Inc.}}
  \bibinfo{year}{2020}\natexlab{c}.
\newblock \bibinfo{title}{{Cloud AutoML}}.
\newblock \bibinfo{howpublished}{\url{https://cloud.google.com/automl}}.
\newblock
\newblock
\shownote{Accessed: 2020-09-01.}


\bibitem[\protect\citeauthoryear{Inc}{Inc}{2020}]%
        {h202020}
\bibfield{author}{\bibinfo{person}{H20.ai Inc}.}
  \bibinfo{year}{2020}\natexlab{}.
\newblock \bibinfo{title}{{H20 Driverless AI}}.
\newblock
  \bibinfo{howpublished}{\url{https://www.h2o.ai/products/h2o-driverless-ai/}}.
\newblock
\newblock
\shownote{Accessed: 2020-09-01.}


\bibitem[\protect\citeauthoryear{Inc.}{Inc.}{2020a}]%
        {ibm2020}
\bibfield{author}{\bibinfo{person}{IBM Inc.}} \bibinfo{year}{2020}\natexlab{a}.
\newblock \bibinfo{title}{{AutoAI with IBM Watson Studio}}.
\newblock
  \bibinfo{howpublished}{\url{https://www.ibm.com/cloud/watson-studio/autoai}}.
\newblock
\newblock
\shownote{Accessed: 2020-09-01.}


\bibitem[\protect\citeauthoryear{Inc.}{Inc.}{2020b}]%
        {azureaoutml2020}
\bibfield{author}{\bibinfo{person}{Microsoft Inc.}}
  \bibinfo{year}{2020}\natexlab{b}.
\newblock \bibinfo{title}{{Azure Machine Learning Studio}}.
\newblock
  \bibinfo{howpublished}{\url{https://azure.microsoft.com/en-us/services/machine-learning/automatedml/}}.
\newblock
\newblock
\shownote{Accessed: 2020-09-01.}


\bibitem[\protect\citeauthoryear{Kandel, Paepcke, Hellerstein, and Heer}{Kandel
  et~al\mbox{.}}{2011}]%
        {2012-profiler}
\bibfield{author}{\bibinfo{person}{Sean Kandel}, \bibinfo{person}{Andreas
  Paepcke}, \bibinfo{person}{Joseph Hellerstein}, {and}
  \bibinfo{person}{Jeffrey Heer}.} \bibinfo{year}{2011}\natexlab{}.
\newblock \showarticletitle{Wrangler: Interactive Visual Specification of Data
  Transformation Scripts}. In \bibinfo{booktitle}{\emph{Proc CHI'11}}.
  \bibinfo{pages}{3363–3372}.
\newblock
\urldef\tempurl%
\url{https://doi.org/10.1145/1978942.1979444}
\showDOI{\tempurl}


\bibitem[\protect\citeauthoryear{Kandel, Paepcke, Hellerstein, and Heer}{Kandel
  et~al\mbox{.}}{2012a}]%
        {Kandel:2012}
\bibfield{author}{\bibinfo{person}{Sean Kandel}, \bibinfo{person}{Andreas
  Paepcke}, \bibinfo{person}{Joseph~M. Hellerstein}, {and}
  \bibinfo{person}{Jeffery Heer}.} \bibinfo{year}{2012}\natexlab{a}.
\newblock \showarticletitle{Enterprise Data Analysis and Visualization: An
  Interview Study}.
\newblock \bibinfo{journal}{\emph{IEEE Transactions on Visualization and
  Computer Graphics}} \bibinfo{volume}{18}, \bibinfo{number}{12}
  (\bibinfo{year}{2012}), \bibinfo{pages}{2917--2926}.
\newblock
\showISSN{2160-9306}
\urldef\tempurl%
\url{https://doi.org/10.1109/TVCG.2012.219}
\showDOI{\tempurl}


\bibitem[\protect\citeauthoryear{Kandel, Parikh, Paepcke, Hellerstein, and
  Heer}{Kandel et~al\mbox{.}}{2012b}]%
        {2011-wrangler}
\bibfield{author}{\bibinfo{person}{Sean Kandel}, \bibinfo{person}{Ravi Parikh},
  \bibinfo{person}{Andreas Paepcke}, \bibinfo{person}{Joseph~M. Hellerstein},
  {and} \bibinfo{person}{Jeffrey Heer}.} \bibinfo{year}{2012}\natexlab{b}.
\newblock \showarticletitle{Profiler: Integrated Statistical Analysis and
  Visualization for Data Quality Assessment}. In \bibinfo{booktitle}{\emph{Proc
  AVI'12}}. \bibinfo{pages}{547–554}.
\newblock
\urldef\tempurl%
\url{https://doi.org/10.1145/2254556.2254659}
\showDOI{\tempurl}


\bibitem[\protect\citeauthoryear{Kim, Zimmermann, DeLine, and Begel}{Kim
  et~al\mbox{.}}{2018}]%
        {Kim:2017}
\bibfield{author}{\bibinfo{person}{Minyung Kim}, \bibinfo{person}{Thomas
  Zimmermann}, \bibinfo{person}{Robert DeLine}, {and} \bibinfo{person}{Andrew
  Begel}.} \bibinfo{year}{2018}\natexlab{}.
\newblock \showarticletitle{Data Scientists in Software Teams: State of the Art
  and Challenges}.
\newblock \bibinfo{journal}{\emph{IEEE Transactions on Software Engineering}}
  \bibinfo{volume}{44}, \bibinfo{number}{11} (\bibinfo{year}{2018}),
  \bibinfo{pages}{1024--1038}.
\newblock
\showISSN{2326-3881}
\urldef\tempurl%
\url{https://doi.org/10.1109/TSE.2017.2754374}
\showDOI{\tempurl}


\bibitem[\protect\citeauthoryear{Lee, Macke, Xin, Lee, Huang, and
  Parameswaran}{Lee et~al\mbox{.}}{2019}]%
        {Lee2019AHP}
\bibfield{author}{\bibinfo{person}{D. Lee}, \bibinfo{person}{Stephen Macke},
  \bibinfo{person}{Doris Xin}, \bibinfo{person}{Angela Lee},
  \bibinfo{person}{Silu Huang}, {and} \bibinfo{person}{Aditya~G.
  Parameswaran}.} \bibinfo{year}{2019}\natexlab{}.
\newblock \showarticletitle{A Human-in-the-loop Perspective on AutoML:
  Milestones and the Road Ahead}.
\newblock \bibinfo{journal}{\emph{IEEE Data Eng. Bull.}} \bibinfo{volume}{42},
  \bibinfo{number}{2} (\bibinfo{year}{2019}), \bibinfo{pages}{59--70}.
\newblock
\urldef\tempurl%
\url{http://sites.computer.org/debull/A19june/p59.pdf}
\showURL{%
\tempurl}


\bibitem[\protect\citeauthoryear{Liao, Gruen, and Miller}{Liao
  et~al\mbox{.}}{2020}]%
        {Liao_2020}
\bibfield{author}{\bibinfo{person}{Q.~Vera Liao}, \bibinfo{person}{Daniel
  Gruen}, {and} \bibinfo{person}{Sarah Miller}.}
  \bibinfo{year}{2020}\natexlab{}.
\newblock \showarticletitle{Questioning the AI: Informing Design Practices for
  Explainable AI User Experiences}.
\newblock \bibinfo{journal}{\emph{Proc CHI'20}} (\bibinfo{year}{2020}),
  \bibinfo{pages}{1–15}.
\newblock
\urldef\tempurl%
\url{https://doi.org/10.1145/3313831.3376590}
\showDOI{\tempurl}


\bibitem[\protect\citeauthoryear{McNutt, Kindlmann, and Correll}{McNutt
  et~al\mbox{.}}{2020}]%
        {McNutt2020}
\bibfield{author}{\bibinfo{person}{Andrew McNutt}, \bibinfo{person}{Gordon
  Kindlmann}, {and} \bibinfo{person}{Michael Correll}.}
  \bibinfo{year}{2020}\natexlab{}.
\newblock \showarticletitle{Surfacing Visualization Mirages}.
\newblock \bibinfo{journal}{\emph{Proc CHI'20}}, \bibinfo{pages}{1–16}.
\newblock
\urldef\tempurl%
\url{https://doi.org/10.1145/3313831.3376420}
\showDOI{\tempurl}


\bibitem[\protect\citeauthoryear{Olson and Kellogg}{Olson and Kellogg}{2014}]%
        {Olson_Kellogg_2014}
\bibfield{author}{\bibinfo{person}{Judith~S. Olson} {and}
  \bibinfo{person}{Wendy~A. Kellogg}.} \bibinfo{year}{2014}\natexlab{}.
\newblock \bibinfo{booktitle}{}.
\newblock \bibinfo{publisher}{Springer New York}.
\newblock
\showISBNx{978-1-4939-0377-1}
\urldef\tempurl%
\url{https://doi.org/10.1007/978-1-4939-0378-8}
\showDOI{\tempurl}


\bibitem[\protect\citeauthoryear{Olson, Bartley, Urbanowicz, and Moore}{Olson
  et~al\mbox{.}}{2016a}]%
        {Olson2019}
\bibfield{author}{\bibinfo{person}{Randal~S. Olson}, \bibinfo{person}{Nathan
  Bartley}, \bibinfo{person}{Ryan~J. Urbanowicz}, {and}
  \bibinfo{person}{Jason~H. Moore}.} \bibinfo{year}{2016}\natexlab{a}.
\newblock \showarticletitle{Evaluation of a Tree-Based Pipeline Optimization
  Tool for Automating Data Science}. In \bibinfo{booktitle}{\emph{Proc GECCO
  '16}}. \bibinfo{pages}{485–492}.
\newblock
\urldef\tempurl%
\url{https://doi.org/10.1145/2908812.2908918}
\showDOI{\tempurl}


\bibitem[\protect\citeauthoryear{Olson, Bartley, Urbanowicz, and Moore}{Olson
  et~al\mbox{.}}{2016b}]%
        {olson2016}
\bibfield{author}{\bibinfo{person}{Randal~S. Olson}, \bibinfo{person}{Nathan
  Bartley}, \bibinfo{person}{Ryan~J. Urbanowicz}, {and}
  \bibinfo{person}{Jason~H. Moore}.} \bibinfo{year}{2016}\natexlab{b}.
\newblock \showarticletitle{Evaluation of a Tree-Based Pipeline Optimization
  Tool for Automating Data Science}. In \bibinfo{booktitle}{\emph{Proc
  GECCO'16}}. \bibinfo{pages}{485–492}.
\newblock
\urldef\tempurl%
\url{https://doi.org/10.1145/2908812.2908918}
\showDOI{\tempurl}


\bibitem[\protect\citeauthoryear{Ono, Castelo, Lopez, Bertini, Freire, and
  Silva}{Ono et~al\mbox{.}}{2020}]%
        {pipelineprofiler2020}
\bibfield{author}{\bibinfo{person}{Jorge~Piazentin Ono}, \bibinfo{person}{Sonia
  Castelo}, \bibinfo{person}{Roque Lopez}, \bibinfo{person}{Enrico Bertini},
  \bibinfo{person}{Juliana Freire}, {and} \bibinfo{person}{Claudio Silva}.}
  \bibinfo{year}{2020}\natexlab{}.
\newblock \showarticletitle{PipelineProfiler: A Visual Analytics Tool for the
  Exploration of AutoML Pipelines}.
\newblock \bibinfo{journal}{\emph{IEEE Transactions on Visualization and
  Computer Graphics}} (\bibinfo{year}{2020}).
\newblock
\urldef\tempurl%
\url{https://doi.org/10.1109/TVCG.2020.3030361}
\showDOI{\tempurl}


\bibitem[\protect\citeauthoryear{Parasuraman, Sheridan, and
  Wickens}{Parasuraman et~al\mbox{.}}{2000}]%
        {Parasuraman2000}
\bibfield{author}{\bibinfo{person}{Raja Parasuraman},
  \bibinfo{person}{Thomas~B. Sheridan}, {and} \bibinfo{person}{Christopher~D.
  Wickens}.} \bibinfo{year}{2000}\natexlab{}.
\newblock \showarticletitle{A Model for Types and Levels of Human Interaction
  with Automation}.
\newblock \bibinfo{journal}{\emph{IEEE Transactions on Systems, Man, and
  Cybernetics - Part A: Systems and Humans}} \bibinfo{volume}{30},
  \bibinfo{number}{3} (\bibinfo{year}{2000}), \bibinfo{pages}{286--297}.
\newblock
\urldef\tempurl%
\url{https://doi.org/10.1109/3468.844354}
\showDOI{\tempurl}


\bibitem[\protect\citeauthoryear{Passi and Jackson}{Passi and Jackson}{2018}]%
        {Passi2018}
\bibfield{author}{\bibinfo{person}{Samir Passi} {and}
  \bibinfo{person}{Steven~J. Jackson}.} \bibinfo{year}{2018}\natexlab{}.
\newblock \showarticletitle{Trust in Data Science: Collaboration, Translation,
  and Accountability in Corporate Data Science Projects}.
\newblock \bibinfo{journal}{\emph{Proc CSCW'2018}} \bibinfo{number}{CSCW},
  Article \bibinfo{articleno}{136} (\bibinfo{year}{2018}),
  \bibinfo{numpages}{28}~pages.
\newblock
\urldef\tempurl%
\url{https://doi.org/10.1145/3274405}
\showDOI{\tempurl}


\bibitem[\protect\citeauthoryear{Pedregosa, Varoquaux, Gramfort, Michel,
  Thirion, Grisel, Blondel, Prettenhofer, Weiss, Dubourg, Vanderplas, Passos,
  Cournapeau, Brucher, Perrot, and {{\'E}}douard Duchesnay}{Pedregosa
  et~al\mbox{.}}{2011}]%
        {scikit-learn}
\bibfield{author}{\bibinfo{person}{Fabian Pedregosa},
  \bibinfo{person}{Ga{{\"e}}l Varoquaux}, \bibinfo{person}{Alexandre Gramfort},
  \bibinfo{person}{Vincent Michel}, \bibinfo{person}{Bertrand Thirion},
  \bibinfo{person}{Olivier Grisel}, \bibinfo{person}{Mathieu Blondel},
  \bibinfo{person}{Peter Prettenhofer}, \bibinfo{person}{Ron Weiss},
  \bibinfo{person}{Vincent Dubourg}, \bibinfo{person}{Jake Vanderplas},
  \bibinfo{person}{Alexandre Passos}, \bibinfo{person}{David Cournapeau},
  \bibinfo{person}{Matthieu Brucher}, \bibinfo{person}{Matthieu Perrot}, {and}
  \bibinfo{person}{{{\'E}}douard Duchesnay}.} \bibinfo{year}{2011}\natexlab{}.
\newblock \showarticletitle{Scikit-learn: Machine Learning in Python}.
\newblock \bibinfo{journal}{\emph{Journal of Machine Learning Research}}
  \bibinfo{volume}{12}, \bibinfo{number}{85} (\bibinfo{year}{2011}),
  \bibinfo{pages}{2825--2830}.
\newblock
\urldef\tempurl%
\url{http://jmlr.org/papers/v12/pedregosa11a.html}
\showURL{%
\tempurl}


\bibitem[\protect\citeauthoryear{Sarikaya, Correll, Bartram, Tory, and
  Fisher}{Sarikaya et~al\mbox{.}}{2019}]%
        {Sarikaya2019}
\bibfield{author}{\bibinfo{person}{Alper Sarikaya}, \bibinfo{person}{Micharl
  Correll}, \bibinfo{person}{Lyn Bartram}, \bibinfo{person}{Melanie Tory},
  {and} \bibinfo{person}{Danyel Fisher}.} \bibinfo{year}{2019}\natexlab{}.
\newblock \showarticletitle{What Do We Talk About When We Talk About
  Dashboards?}
\newblock \bibinfo{journal}{\emph{IEEE Transactions on Visualization and
  Computer Graphics}} \bibinfo{volume}{25}, \bibinfo{number}{1}
  (\bibinfo{year}{2019}), \bibinfo{pages}{682--692}.
\newblock
\urldef\tempurl%
\url{https://doi.org/10.1109/TVCG.2018.2864903}
\showDOI{\tempurl}


\bibitem[\protect\citeauthoryear{Taska, Miller, Hughes, Markow, and
  Braganza}{Taska et~al\mbox{.}}{2017}]%
        {datasci_2017}
\bibfield{author}{\bibinfo{person}{Bledi Taska}, \bibinfo{person}{Steven~M.
  Miller}, \bibinfo{person}{Debbie Hughes}, \bibinfo{person}{Will Markow},
  {and} \bibinfo{person}{Soumya Braganza}.} \bibinfo{year}{2017}\natexlab{}.
\newblock \bibinfo{title}{The Quant Crunch: How the Demand for Data Science
  Skills is Disrupting the Job Market}.
\newblock
\newblock
\urldef\tempurl%
\url{https://www.ibm.com/downloads/cas/3RL3VXGA}
\showURL{%
\tempurl}


\bibitem[\protect\citeauthoryear{Wang, Weisz, Muller, Ram, Geyer, Dugan,
  Tausczik, Samulowitz, and Gray}{Wang et~al\mbox{.}}{2019b}]%
        {Wang:2019}
\bibfield{author}{\bibinfo{person}{Dakuo Wang}, \bibinfo{person}{Justin~D.
  Weisz}, \bibinfo{person}{Michael Muller}, \bibinfo{person}{Parikshit Ram},
  \bibinfo{person}{Werner Geyer}, \bibinfo{person}{Casey Dugan},
  \bibinfo{person}{Yla Tausczik}, \bibinfo{person}{Horst Samulowitz}, {and}
  \bibinfo{person}{Alexander Gray}.} \bibinfo{year}{2019}\natexlab{b}.
\newblock \showarticletitle{Human-AI Collaboration in Data Science: Exploring
  Data Scientists’ Perceptions of Automated AI}.
\newblock \bibinfo{journal}{\emph{Proc. CSCW'19}}, 24.
\newblock
\urldef\tempurl%
\url{https://doi.org/10.1145/3359313}
\showDOI{\tempurl}


\bibitem[\protect\citeauthoryear{Wang, Ming, Jin, Shen, Liu, Smith,
  Veeramachaneni, and Qu}{Wang et~al\mbox{.}}{2019a}]%
        {wang2018atmseer}
\bibfield{author}{\bibinfo{person}{Qianwen Wang}, \bibinfo{person}{Yao Ming},
  \bibinfo{person}{Zhihua Jin}, \bibinfo{person}{Qiaomu Shen},
  \bibinfo{person}{Dongyu Liu}, \bibinfo{person}{Micah~J. Smith},
  \bibinfo{person}{Kalyan Veeramachaneni}, {and} \bibinfo{person}{Huamin Qu}.}
  \bibinfo{year}{2019}\natexlab{a}.
\newblock \showarticletitle{ATMSeer: Increasing Transparency and
  Controllability in Automated Machine Learning}. In
  \bibinfo{booktitle}{\emph{Proc CHI'19}}. \bibinfo{pages}{1–12}.
\newblock
\urldef\tempurl%
\url{https://doi.org/10.1145/3290605.3300911}
\showDOI{\tempurl}


\bibitem[\protect\citeauthoryear{Weidele, Weisz, Oduor, Muller, Andres, Gray,
  and Wang}{Weidele et~al\mbox{.}}{2020}]%
        {Weidele_2020}
\bibfield{author}{\bibinfo{person}{Daniel Karl~I. Weidele},
  \bibinfo{person}{Justin~D. Weisz}, \bibinfo{person}{Erick Oduor},
  \bibinfo{person}{Michael Muller}, \bibinfo{person}{Josh Andres},
  \bibinfo{person}{Alexander Gray}, {and} \bibinfo{person}{Dakuo Wang}.}
  \bibinfo{year}{2020}\natexlab{}.
\newblock \showarticletitle{AutoAIViz: Opening the Blackbox of Automated
  Artificial Intelligence with Conditional Parallel Coordinates}. In
  \bibinfo{booktitle}{\emph{Proc IUI'20}}. \bibinfo{pages}{308–312}.
\newblock
\urldef\tempurl%
\url{https://doi.org/10.1145/3377325.3377538}
\showDOI{\tempurl}


\bibitem[\protect\citeauthoryear{Wongsuphasawat, Smilkov, Wexler, Wilson,
  Man\'{e}, Fritz, Krishnan, Viégas, and Wattenberg}{Wongsuphasawat
  et~al\mbox{.}}{2018}]%
        {Ham:2018}
\bibfield{author}{\bibinfo{person}{Kanit Wongsuphasawat},
  \bibinfo{person}{Daniel Smilkov}, \bibinfo{person}{James Wexler},
  \bibinfo{person}{Jimbo Wilson}, \bibinfo{person}{Dandelion Man\'{e}},
  \bibinfo{person}{Doug Fritz}, \bibinfo{person}{Dilip Krishnan},
  \bibinfo{person}{Fernanda~B. Viégas}, {and} \bibinfo{person}{Martin
  Wattenberg}.} \bibinfo{year}{2018}\natexlab{}.
\newblock \showarticletitle{Visualizing Dataflow Graphs of Deep Learning Models
  in TensorFlow}.
\newblock \bibinfo{journal}{\emph{IEEE Transactions on Visualization and
  Computer Graphics}} \bibinfo{volume}{24}, \bibinfo{number}{1}
  (\bibinfo{year}{2018}), \bibinfo{pages}{1--12}.
\newblock
\urldef\tempurl%
\url{https://doi.org/10.1109/TVCG.2017.2744878}
\showDOI{\tempurl}


\bibitem[\protect\citeauthoryear{Yao, Wang, Escalante, Guyon, Hu, Li, Tu, Yang,
  and Yu}{Yao et~al\mbox{.}}{2018}]%
        {yao2018}
\bibfield{author}{\bibinfo{person}{Quanming Yao}, \bibinfo{person}{Mengshuo
  Wang}, \bibinfo{person}{Hugo~Jair Escalante}, \bibinfo{person}{Isabelle
  Guyon}, \bibinfo{person}{Yi{-}Qi Hu}, \bibinfo{person}{Yu{-}Feng Li},
  \bibinfo{person}{Wei{-}Wei Tu}, \bibinfo{person}{Qiang Yang}, {and}
  \bibinfo{person}{Yang Yu}.} \bibinfo{year}{2018}\natexlab{}.
\newblock \bibinfo{title}{Taking Human out of Learning Applications: A Survey
  on Automated Machine Learning}.
\newblock
\newblock
\urldef\tempurl%
\url{http://arxiv.org/abs/1810.13306}
\showURL{%
\tempurl}


\bibitem[\protect\citeauthoryear{Yuan, Chen, Yang, Liu, Xia, and Liu}{Yuan
  et~al\mbox{.}}{2020}]%
        {yuan2020survey}
\bibfield{author}{\bibinfo{person}{Jun Yuan}, \bibinfo{person}{Changjian Chen},
  \bibinfo{person}{Weikai Yang}, \bibinfo{person}{Mengchen Liu},
  \bibinfo{person}{Jiazhi Xia}, {and} \bibinfo{person}{Shixia Liu}.}
  \bibinfo{year}{2020}\natexlab{}.
\newblock \showarticletitle{A Survey of Visual Analytics Techniques for Machine
  Learning}.
\newblock \bibinfo{journal}{\emph{Compunational Visual Media}}
  (\bibinfo{year}{2020}).
\newblock
\urldef\tempurl%
\url{https://doi.org/10.1007/s41095-020-0191-7}
\showDOI{\tempurl}


\bibitem[\protect\citeauthoryear{Zhang, Muller, and Wang}{Zhang
  et~al\mbox{.}}{2020b}]%
        {zhang:2020}
\bibfield{author}{\bibinfo{person}{Amy~X. Zhang}, \bibinfo{person}{Michael
  Muller}, {and} \bibinfo{person}{Dakuo Wang}.}
  \bibinfo{year}{2020}\natexlab{b}.
\newblock \showarticletitle{How Do Data Science Workers Collaborate? Roles,
  Workflows, and Tools}.
\newblock \bibinfo{journal}{\emph{Proc CSCW'2020}}, Article
  \bibinfo{articleno}{022} (\bibinfo{date}{May} \bibinfo{year}{2020}),
  \bibinfo{numpages}{23}~pages.
\newblock
\urldef\tempurl%
\url{https://doi.org/10.1145/3392826}
\showDOI{\tempurl}


\bibitem[\protect\citeauthoryear{Zhang, Liao, and Bellamy}{Zhang
  et~al\mbox{.}}{2020a}]%
        {ZhangY2020}
\bibfield{author}{\bibinfo{person}{Yunfeng Zhang}, \bibinfo{person}{Q.~Vera
  Liao}, {and} \bibinfo{person}{Rachel K.~E. Bellamy}.}
  \bibinfo{year}{2020}\natexlab{a}.
\newblock \showarticletitle{Effect of Confidence and Explanation on Accuracy
  and Trust Calibration in AI-Assisted Decision Making}. In
  \bibinfo{booktitle}{\emph{Proc FAccT*'20}}. \bibinfo{pages}{295–305}.
\newblock
\urldef\tempurl%
\url{https://doi.org/10.1145/3351095.3372852}
\showDOI{\tempurl}


\bibitem[\protect\citeauthoryear{Z\"{o}ller and Huber}{Z\"{o}ller and
  Huber}{2019}]%
        {zoller2019benchmark}
\bibfield{author}{\bibinfo{person}{Marc-André Z\"{o}ller} {and}
  \bibinfo{person}{Marco~F. Huber}.} \bibinfo{year}{2019}\natexlab{}.
\newblock \bibinfo{title}{Benchmark and Survey of Automated Machine Learning
  Frameworks}.
\newblock
\newblock
\urldef\tempurl%
\url{https://arxiv.org/abs/1904.12054}
\showURL{%
\tempurl}


\end{thebibliography}




\end{document}